\documentclass[twocolumn,tighten]{aastex63}

\usepackage{natbib}
\usepackage{multirow}
\usepackage{amsmath}
\usepackage{hyperref}
\hypersetup{citecolor=blue,urlcolor=blue}
\received{May 18, 2020}
\revised{June 10, 2020}
\accepted{June 11, 2020}
\submitjournal{ApJ}

\shorttitle{Enhanced [CI](1--0) emission in the type-1 Seyfert galaxy NGC 7469}
\shortauthors{T. Izumi et al.}
\begin{document}
\title{ALMA Observations of Multiple-CO and C Lines Toward the Active Galactic Nucleus of NGC 7469:\\ 
X-Ray-dominated Region Caught in the Act}

\correspondingauthor{Takuma Izumi}
\email{takuma.izumi@nao.ac.jp}

\author[0000-0002-0786-7307]{Takuma Izumi}
\altaffiliation{NAOJ Fellow}
\affil{National Astronomical Observatory of Japan, 2-21-1 Osawa, Mitaka, Tokyo 181-8588, Japan}
\affil{Department of Astronomical Science, The Graduate University for Advanced Studies, SOKENDAI, 2-21-1 Osawa, Mitaka, Tokyo 181-8588, Japan}
\author{Dieu D. Nguyen}
\affil{National Astronomical Observatory of Japan, 2-21-1 Osawa, Mitaka, Tokyo 181-8588, Japan}
\author{Masatoshi Imanishi}
\affil{National Astronomical Observatory of Japan, 2-21-1 Osawa, Mitaka, Tokyo 181-8588, Japan}
\affil{Department of Astronomical Science, The Graduate University for Advanced Studies, SOKENDAI, 2-21-1 Osawa, Mitaka, Tokyo 181-8588, Japan}
\author{Taiki Kawamuro}
\altaffiliation{JSPS Fellow}
\affil{National Astronomical Observatory of Japan, 2-21-1 Osawa, Mitaka, Tokyo 181-8588, Japan}
\author{Shunsuke Baba}
\altaffiliation{JSPS Fellow}
\affil{National Astronomical Observatory of Japan, 2-21-1 Osawa, Mitaka, Tokyo 181-8588, Japan}
\author{Suzuka Nakano}
\affil{Department of Astronomical Science, The Graduate University for Advanced Studies, SOKENDAI, 2-21-1 Osawa, Mitaka, Tokyo 181-8588, Japan}
\author{Kotaro Kohno}
\affil{Institute of Astronomy, Graduate School of Science, The University of Tokyo, 2-21-1 Osawa, Mitaka, Tokyo 181-0015, Japan}
\affil{Research Center for the Early Universe, Graduate School of Science, The University of Tokyo, 7-3-1 Hongo, Bunkyo, Tokyo 113-0033, Japan}
\author{Satoki Matsushita}
\affil{Institute of Astronomy and Astrophysics, Academia Sinica 11F of Astronomy-Mathematics Building, AS/NTU, No.1, Sec.4, Roosevelt Rd, Taipei 10617, Taiwan, R.O.C.}
\author{David S. Meier}
\affil{Department of Physics, New Mexico Institute of Mining and Technology, Socorro, NM 87801, USA}
\author{Jean L. Turner}
\affil{UCLA Department of Physics and Astronomy, Los Angeles, CA 90095-1547, USA}
\author{Tomonari Michiyama}
\affil{Kavli Institute for Astronomy and Astrophysics, Peking University, 5 Yiheyuan Road, Haidian District, Beijing 100871, P.R.China}
\author{Nanase Harada}
\affil{Institute of Astronomy and Astrophysics, Academia Sinica 11F of Astronomy-Mathematics Building, AS/NTU, No.1, Sec.4, Roosevelt Rd, Taipei 10617, Taiwan, R.O.C.}
\author{Sergio Mart\'{i}n}
\affil{European Southern Observatory, Alonso de C\'ordova 3107, Vitacura, Santiago, 763-0355 Chile}
\affil{Joint ALMA Observatory, Alonso de C\'ordova, 3107, Vitacura, Santiago 763-0355, Chile}
\author{Kouichiro Nakanishi}
\affil{National Astronomical Observatory of Japan, 2-21-1 Osawa, Mitaka, Tokyo 181-8588, Japan}
\affil{Department of Astronomical Science, The Graduate University for Advanced Studies, SOKENDAI, 2-21-1 Osawa, Mitaka, Tokyo 181-8588, Japan}
\author{Shuro Takano}
\affil{Department of Physics, General Studies, College of Engineering, Nihon University, Tamuramachi, Koriyama, Fukushima 963-8642, Japan}
\author{Tommy Wiklind}
\affil{Catholic University of America, Department of Physics, Washington, DC 20064, USA}
\author{Naomasa Nakai}
\affil{School of Science and Technology, Kwansei Gakuin University, 2-1 Gakuen, Sanda, Hyogo 669-1337, Japan}
\affil{Tomonaga Center for the History of the Universe, University of Tsukuba, Tsukuba, Ibaraki 305-8571, Japan}
\author{Pei-Ying Hsieh}
\affil{Institute of Astronomy and Astrophysics, Academia Sinica 11F of Astronomy-Mathematics Building, AS/NTU, No.1, Sec.4, Roosevelt Rd., Taipei 10617, Taiwan, R.O.C.}

\begin{abstract}
We used the Atacama Large Millimeter/submillimeter Array (ALMA) to map 
$^{12}$CO($J$ = 1--0), $^{12}$CO($J$ = 2--1), $^{12}$CO($J$ = 3--2), 
$^{13}$CO($J$ = 2--1), and [\ion{C}{1}]($^3P_1$--$^3P_0$) emission lines 
around the type 1 active galactic nucleus (AGN) of NGC 7469 ($z = 0.0164$) at $\sim 100$ pc resolutions. 
The CO lines are bright in both the circumnuclear disk (central $\sim 300$ pc) 
and the surrounding starburst (SB) ring ($\sim 1$ kpc diameter), 
with two bright peaks on either side of the AGN. 
By contrast, the [\ion{C}{1}]($^3P_1$--$^3P_0$) line is strongly peaked on the AGN. 
Consequently, the brightness temperature ratio of [\ion{C}{1}]($^3P_1$--$^3P_0$) 
to $^{13}$CO(2--1) is $\sim 20$ at the AGN, as compared to $\sim 2$ in the SB ring. 
Our local thermodynamic equilibrium (LTE) and non-LTE models indicate that 
the enhanced line ratios (or \ion{C}{1} enhancement) are due to an elevated C$^0$/CO abundance ratio ($\sim 3-10$) 
and temperature ($\sim 100-500$ K) around the AGN as compared to the SB ring (abundance ratio $\sim 1$, temperature $\lesssim 100$ K), 
which accords with the picture of the X-ray-dominated Region (XDR). 
Based on dynamical modelings, we also provide CO(1--0)-to- and [\ion{C}{1}]($^3P_1$--$^3P_0$)-to-molecular mass 
conversion factors at the central $\sim 100$ pc of this AGN as $\alpha_{\rm CO} = 4.1$ 
and $\alpha_{\rm CI} = 4.4~M_\odot$ (K km s$^{-1}$ pc$^2$)$^{-1}$, respectively. 
Our results suggest that the \ion{C}{1} enhancement is potentially a good marker of AGNs 
that could be used in a new submillimeter diagnostic method toward dusty environments. 
\end{abstract}

\section{Introduction}\label{sec1}
The cold gas, particularly in the molecular phase in the centers of galaxies, 
plays a key role in the evolution of galaxies because it is the site of massive star formation 
as well as the reservoir of fuel for central supermassive black holes (SMBHs). 
The mass accretion onto a SMBH produces enormous amounts of energy observable as 
an active galactic nucleus (AGN), which is much more efficient 
in producing X-ray radiation than massive stars \citep{2018ARA&A..56..625H}. 
Consequently, we would expect that such a different heating mechanism will produce different signatures 
on the circumnuclear gas properties \citep{2005A&A...436..397M,2007A&A...461..793M}. 
for example, photodissociation regions (PDRs) caused by intense ultra-violet (UV) 
radiation from massive stars \citep[e.g.,][]{1997ARA&A..35..179H,1999RvMP...71..173H} 
likely give way to X-ray dominated regions (XDRs), 
where gas physical and chemical properties are governed 
by the harsh X-ray irradiation from the central AGN \citep[e.g.,][]{1996A&A...306L..21L,1996ApJ...466..561M}. 
Cosmic rays from supernovae and the injection of mechanical energy 
induced by AGN jet/outflow also make unique chemical compositions 
\citep[e.g.,][]{2011A&A...525A.119M,2012A&A...542A..65K,2015A&A...574A.127K}. 

Diagnosing these energy sources by sub/millimeter spectroscopic observations can be 
useful to uncover dust-obscured activity because these wavelengths do not suffer from severe dust extinction. 
For example, so-called obscured (total obscuring column $N_{\rm H} \gtrsim 10^{23}$ cm$^{-2}$) 
AGNs account for $\gtrsim 50\%$ of the total AGN population at least at $z \sim 0-2$ \citep{2015ApJ...802...89B}. 
Hence, a robust mm/submm energy diagnostics is quite beneficial to obtain a comprehensive picture on, e.g., the cosmic evolution of SMBHs. 

Based on these interests, many key molecules have been suggested as useful observational diagnostic tools. 
Among them, an enhanced HCN intensity relative to those of CO, HCO$^+$, or CS 
\citep[e.g.,][]{1994ApJ...426L..77T,2001ASPC..249..672K,2005AIPC..783..203K,
2008ApJ...677..262K,2013PASJ...65..100I,2007AJ....134.2366I,2016AJ....152..218I} 
may be a unique feature to AGNs. 
Extensive modelings of the observed line ratios suggest that an enhancement of HCN abundance 
would be a key to explain the intensity enhancement \citep{2016ApJ...818...42I}, 
which accords with the X-ray-induced chemistry \citep{1996A&A...306L..21L,1996ApJ...466..561M} 
or with the chemistry rather generally expected in high temperature regions \citep{2010ApJ...721.1570H,2013ApJ...765..108H}, 
although there are counter-arguments for the reliability of this HCN-enhancement in AGNs \citep[e.g.,][]{2011A&A...528A..30C,2020ApJ...893..149P}. 
One difficulty of studying this HCN-enhancement is the interpretation of the ratio from the perspectives of 
line excitation, opacity, and particularly the complex (time-dependent) chemistry as discussed in \citet{2016ApJ...818...42I}. 
Maser amplification of HCN intensity due to infrared-pumping may also matter \citep{2015ApJ...799...26M}. 
Hence, while keeping further investigation on the robust origin of the HCN-enhancement, 
another effort to explore alternative, and simpler to interpret if possible, sub/mm diagnostic methods is valuable. 

Given this situation, here we focus on the submm atomic carbon emission line [\ion{C}{1}]($^3P_1$--$^3P_0$) 
(denoted as [\ion{C}{1}](1--0) hereafter) at the rest frequency of $\nu_{\rm rest}$ = 492.1607 GHz. 
Because of the high abundance of C$^0$ atom and low energies of its fine structure levels from the ground state, 
the two lines of the [\ion{C}{1}] triplet are important coolant of the neutral interstellar medium (ISM). 
As C$^0$ is a very fundamental form of the carbon-bearing species, 
related formation and destruction processes are relatively simple to understand. 
In a classical PDR scheme, C$^0$ atoms distribute in a thin layer between 
a fully ionized \ion{H}{2} region and a molecular core \citep{1997ARA&A..35..179H,1999RvMP...71..173H}. 
More recent refinements reached a general conclusion that C$^0$ 
rather co-exists with molecular CO due to, e.g., turbulent mixing \citep{2015MNRAS.448.1607G}, 
non-equilibrium chemistry \citep{1997A&A...323L..13S}, 
influence of cosmic rays \citep{2004MNRAS.351..147P,2018MNRAS.478.1716P}, 
or clumpiness \citep{1993ApJ...405..216M}. 

This global spatial concomitance has been confirmed in Galactic star-forming regions 
\citep[e.g.,][]{1997IAUS..178..129K,2000ApJ...539L.133P,1999ApJ...527L..59I,2002ApJS..139..467I,2013ApJ...774L..20S}. 
Although PDR surfaces cannot be resolved, such similar spatial distributions 
are also found in nearby starburst galaxies \citep{2016A&A...592L...3K,2019ApJ...887..143S}. 
These consequently constitute the basis for the use of [\ion{C}{1}](1--0) 
as a potential tracer of H$_2$ mass, and the line is now 
enthusiastically observed both in nearby \citep[e.g.,][]{2004ApJ...615L..29P,2012ApJ...753...70K,
2015A&A...578A..95I,2017ApJ...840L..18J,2019ApJ...887..105C,2019ApJ...880..133J} 
and high-redshift galaxies \citep[e.g.,][]{2004MNRAS.351..147P,2011ApJ...730...18W,2013MNRAS.435.1493A,
2017MNRAS.466.2825B,2017A&A...602A..11P,2018ApJ...869...27V,2020ApJ...890...24V,2019A&A...624A..23N,2020ApJ...889L...7H}. 
Note that however, there is a distinct difference in, for example [\ion{C}{1}](1--0)/$^{12}$CO(1--0) line ratio 
between Galactic star-forming clouds and starburst galaxies. 
Central region of nearby starburst galaxies show [\ion{C}{1}](1--0)/$^{12}$CO(1--0) 
$\sim 0.1-0.3$ \citep[brightness temperature $T_{\rm B}$ unit;][]{2000ApJ...537..644G,2016A&A...592L...3K}, 
whereas it is usually $\lesssim 0.1$ in Milky Way objects \citep[e.g.,][]{1991ApJ...381..200W,2001ApJ...548..253O,2005ApJ...623..889O} 
except for the harsh central molecular zone (CMZ) with $\sim 0.1-0.3$ \citep{2001ApJ...548..253O}. 
Hence pertaining physical or chemical conditions would be 
different among these Galactic clouds and starburst galaxies. 

As compared to PDRs, the higher column-penetrating power 
and higher dissociating/ionizing nature of X-rays 
will more efficiently enhance C$^0$ abundance in XDRs relative to CO 
over a larger volume of molecular cloud as the CO molecule is easily dissociated therein 
\citep{2005A&A...436..397M,2007A&A...461..793M}. 
Although most of the previous [\ion{C}{1}](1--0) observations toward extragalactic objects 
were conducted with ground-based single-dish telescopes or {\it Herschel} Space Observatory 
that mixed up $\gtrsim$a few kpc scale emission \citep[e.g.,][]{2000ApJ...537..644G,2015A&A...578A..95I,2018ApJ...869...27V}, 
\citet{2002A&A...383...82I} showed that the ratio of [\ion{C}{1}](1--0) to $^{13}$CO(2--1), 
both of which would be optically thin, depends on a type of a galaxy: 
it is lowest for quiescent galaxies ($\lesssim 0.5-1$), moderate in starburst galaxies ($\sim 1-3$), 
and highest for AGN-host galaxies ($\sim 4-5$). 
The high [\ion{C}{1}](1--0)/$^{13}$CO(2--1) ratio of $\sim 1-5$ ($T_{\rm B}$ unit), 
which is usually not seen in PDRs of our Galaxy \citep{1997IAUS..178..129K}, 
is at least partly attributed to elevated C$^0$ abundances in extragalactic nuclei via radiative transfer calculations. 
More recently, \citet{2018ApJ...867...48I} found in the Circinus galaxy that [\ion{C}{1}](1--0)/CO(3--2) flux ratio 
increases significantly when it is measured closer to the AGN position based on ALMA observations, 
indicating an influence of the AGN on the nature of the surrounding cold gas. 

It is remarkable in this context that one-zone PDR and XDR models of \citet{2005A&A...436..397M} 
show that the [\ion{C}{1}](1--0)/$^{13}$CO(2--1) ratio can be $>10\times$ higher 
in XDRs than in PDRs \citep{2007A&A...461..793M}, 
due to the more efficient CO dissociation and more harsh excitation conditions in the former. 
However, such a distinct difference between AGNs and starburst galaxies 
was not observed in the single dish observations of \citet{2002A&A...383...82I}. 
This could be due to their insufficient resolution to isolate XDRs, 
which have a characteristic size (diameter) of $\lesssim 100$ pc 
or $\lesssim 1\arcsec$ in the nearby universe \citep{2010A&A...513A...7S}. 
We thus need high resolution comprehensive observations 
of C$^0$ and CO lines to faithfully measure line flux (or abundance) 
ratios in XDRs, which then give a robust basis for the [\ion{C}{1}](1--0)-based diagnostics. 
Now this can be accomplished by the Atacama Large Millimeter/submillimeter Array (ALMA), 
which provides unprecedented high angular resolutions and sensitivities.

\subsection{The Target Galaxy NGC 7469} 
\begin{figure}
\begin{center}
\includegraphics[width=\linewidth]{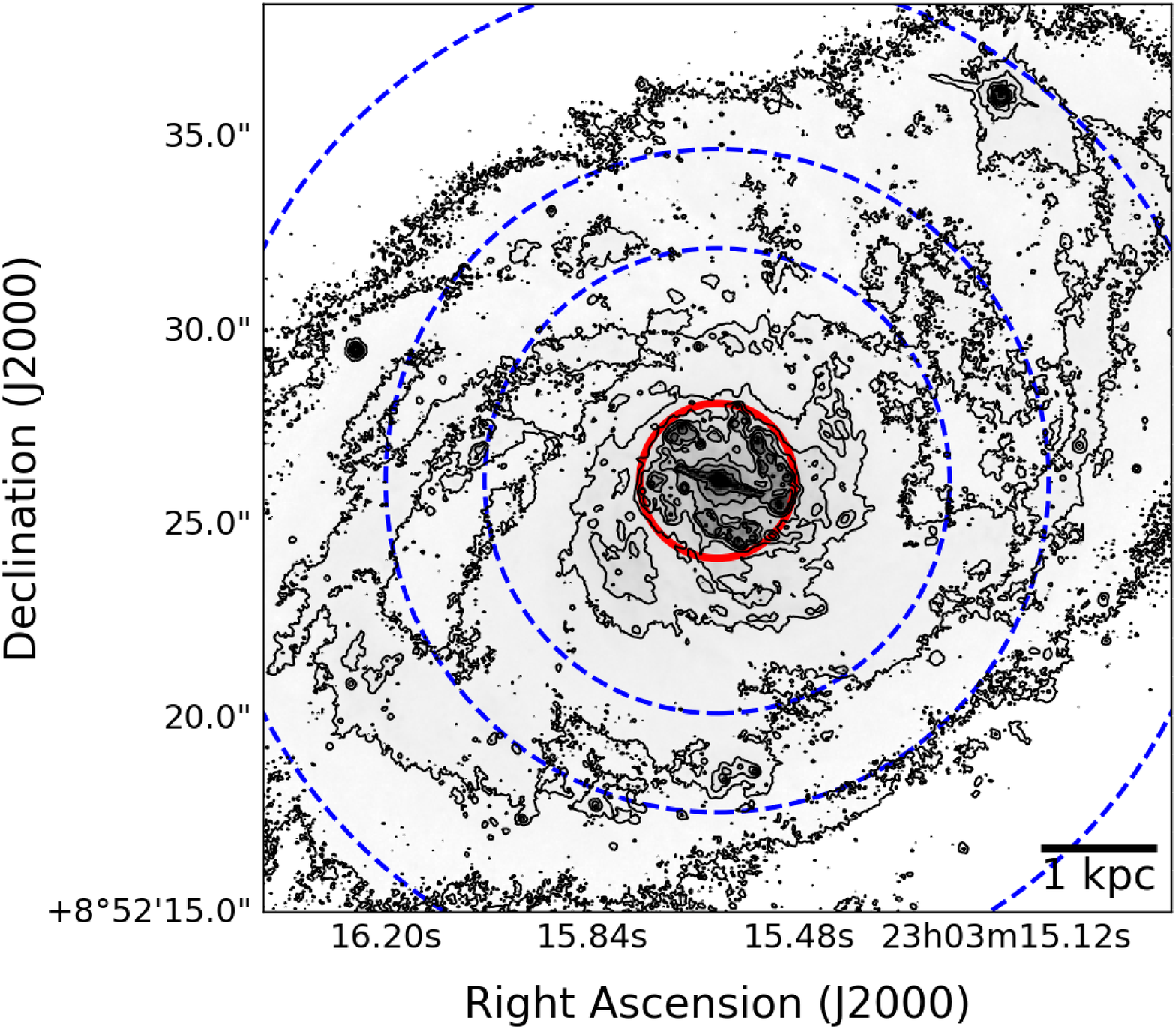}
\caption{
A $B$-band image of NGC 7469 acquired by 
the {\it Hubble Space Telescope (HST)} Advanced Camera for Surveys (ACS)/F435W. 
The gray scale and the contours only indicate counts. 
We extracted this data from the {\it HST} Legacy Archive. 
The regions of our interest, starburst ring and the circumnuclear disk, 
are encompassed by the red circle ($r = 2\arcsec$). 
The field of views (FoVs) of our ALMA observations are indicated 
by the blue dashed circles (smallest one = Band 8 FoV, then becomes larger for Band 7, 6$...$). 
Note that the central pixel (AGN) is saturated, hence the bar-like structure inside the red circle is an artifact. 
}
\label{fig1}
\end{center}
\end{figure}

In this work we present high resolution ALMA observations 
of multiple CO and [\ion{C}{1}](1--0) lines toward NGC 7469 (Figure \ref{fig1}) 
to study detailed line emission distributions and flux ratios, 
with our particular attention on the [\ion{C}{1}](1--0)/$^{13}$CO(2--1) ratio. 
NGC 7469 is an active barred spiral galaxy located at $D = 71.2$ Mpc ($z = 0.01641$ and 1$\arcsec$ = 334 pc)
\footnote{Based on the NASA/IPAC Extragalactic Database (NED). A standard cosmology of $H_0$ = 70 km s$^{-1}$ Mpc$^{-1}$, 
$\Omega_M$ = 0.3, and $\Omega_\Lambda$ = 0.7 is assumed throughout this paper.}. 
It hosts a luminous type-1 Seyfert nucleus with an absorption-corrected 2--10 keV luminosity of 
$L_{\rm 2-10keV} = 1.5 \times 10^{43}$ erg s$^{-1}$ \citep{2014ApJ...783..106L}, 
as evidenced by broad Balmer emission lines \citep{2014ApJ...795..149P}. 
The time-variability in UV to X-ray bands \citep{2000ApJ...544..734N,2000ApJ...535...58K}, 
as well as fast ionized outflows emanating from the nucleus \citep{2007A&A...466..107B,2020arXiv200204031C} 
confirm the genuine existence of AGN in this galaxy. 
NGC 7469 is also categorized as luminous infrared galaxy (LIRG) owing to its high IR luminosity \citep[$L_{\rm 8-1000\mu m} = 10^{11.6}~L_\odot$,][]{2003AJ....126.1607S}. 
There is a $\sim 300$ pc diameter circumnuclear disk (CND) at the center \citep[e.g.,][]{2004ApJ...602..148D,2015ApJ...811...39I}, 
which is surrounded by a luminous starburst ring with a radius of $\sim 500$ pc \citep[e.g.,][]{2003AJ....126..143S,2007ApJ...661..149D}. 
Relevant properties of NGC 7469 are summarized in Table \ref{tbl1}. 

In the CND, multi-wavelength observations at centimeter \citep{2003ApJ...592..804L}, 
$K$-band \citep{1995ApJ...444..129G}, and 3.3 $\mu$m 
polycyclic aromatic hydrocarbon feature \citep{2004ApJ...617..214I,2014ApJ...780...86E}, 
all indicate that the AGN is energetically dominant. 
Reverberation mapping observations revealed that the mass of 
the central SMBH is $M_{\rm BH} \sim 1 \times 10^7~M_\odot$ \citep{2014ApJ...795..149P}, 
with an Eddington ratio of $\sim 0.3$ \citep{2004A&A...413..477P}. 
The starburst ring is incredibly bright at various wavelengths 
including centimeter \citep{1991ApJ...381...79W,2010MNRAS.401.2599O}, 
submm \citep{2015ApJ...811...39I,2016AJ....152..218I}, 
far-infrared \citep[FIR,][]{2000ApJ...537..631P}, 
mid-infrared \citep[MIR,][]{2003AJ....126..143S,2007ApJ...661..149D}, 
near-infrared \citep[NIR,][]{1995ApJ...444..129G,2000AJ....119..991S}, 
and optical to UV \citep{1998ApJS..117...25M,2000MNRAS.311..120D,2007A&A...467..559C}, 
with an area-integrated star-formation rate (SFR) of 
as high as $\sim 30-50~M_\odot$ yr$^{-1}$ \citep{1995ApJ...444..129G,2011A&A...535A..93P}. 
By jointly analyzing the multi-wavelength data, \citet{2007ApJ...661..149D} revealed 
that many dusty, young ($< 100$ Myr), and massive star clusters 
(individual mass $\sim 10^{6-7}~M_\odot$) are embedded in the ring. 
Furthermore, low-$J$ CO observations revealed the existence of a large amount of cold molecular gas ($>10^9~M_\odot$) 
in the central $\lesssim 2$ kpc region \citep{1990ApJ...354..158M,2004ApJ...602..148D}. 
We have secured efficient sub/mm observations toward this object thus far, 
indicating abundant molecular gas \citep{2015ApJ...811...39I,2016ApJ...827...81I,2015ApJ...806L..34F}. 
Therefore, NGC 7469 provides an optimal site to simultaneously investigate how AGN 
and starburst influence their surrounding gas when observed at high resolutions, 
and may also serve as a local template to better understand the ISM properties of high redshift quasars. 

\begin{deluxetable}{ccc}
\tabletypesize{\small}
\tablecaption{Properties of NGC 7469\label{tbl1}}
\tablewidth{0pt}
\tablehead{
\colhead{Parameter} & \colhead{Value} & \colhead{Ref.}
}
\decimalcolnumbers
\startdata
RC3 morphology & (R$'$)SAB(rs)a & (1) \\
Position of the nucleus &  & (2) \\
$\alpha_{\rm ICRS}$ & 23$^{\rm h}$03$^{\rm m}$15$^{\rm s}$.617 &  \\
$\delta_{\rm ICRS}$ & +08$\arcdeg$52$\arcmin$26$\arcsec$.00 &  \\
Position angle [${}^\circ$] & 128 & (3) \\
Inclination angle [${}^\circ$] & 45 & (3) \\
Systemic Velocity [km s$^{-1}$] & 4920 ($z$=0.01641) & (2) \\
Distance [Mpc] & 71.2 &  \\
Linear scale [pc arcsec$^{-1}$] & 334 &  \\
Nuclear activity & Seyfert 1 & (4) \\
$L_{\rm 2-10keV}$ [erg s$^{-1}$] & 1.5 $\times$ 10$^{43}$ & (5) \\
$L_{\rm IR}$ [$L_\odot$] & 4 $\times$ 10$^{11}$ & (6) \\
$<$SFR$>$ (CND) [$M_\odot$ yr$^{-1}$ kpc$^{-2}$] & 50--100 & (7) \\
Stellar age (CND) [Myr] & 110--190 & (7) \\
\enddata
\tablecomments{The $<$SFR$>$ (CND) indicates the averaged star formation rate 
over the central $\sim$ 1$''$ region (i.e., circumnuclear disk = CND). 
The stellar age is also measured for the CND. 
Kinematic parameters are derived based on CO observations. 
(1) \citet{1991rc3..book.....D}; (2) This work; (3) \citet{2004ApJ...602..148D}; (4) \citet{1993ApJ...414..552O}; 
(5) \citet{2014ApJ...783..106L}; (6) \citet{2003AJ....126.1607S}; (7) \citet{2007ApJ...671.1388D}.}
\end{deluxetable}

The structure of this paper is as follows. 
In \S~2, we describe our ALMA observations. 
\S~3 show the results of our observations including 
spatial distributions of the line emission, line profiles, and line ratios. 
We will discuss possible origins of enhanced ratios of [\ion{C}{1}](1--0) to CO lines in \S~4. 
A conversion factor from [\ion{C}{1}](1--0)-to- and CO(1--0)-to-H$_2$ mass in the CND is also presented there. 
Our conclusions of this paper are summarized in \S~5.

\section{Observations and data reduction}\label{sec2} 
Our aim of this program is to investigate detailed feedback of AGN and starburst activity on their surrounding medium. 
We observed multi-phase gas lines including 
the molecular lines $^{12}$CO(1--0), (2--1), (3--2), 
an optically thin isotopologue $^{13}$CO(2--1), 
and the atomic carbon [\ion{C}{1}](1--0) line, 
as well as their underlying continuum emission by using ALMA. 
The relevant excitation parameters of these lines can be found in Table \ref{tbl2}. 
Our observations were conducted during Cycle 5 (project ID: \#2017.1.00078.S, PI = T. Izumi) 
using the Band 3, 6, 7, and 8 receivers, from 2017 December to 2018 September. 
The phase-tracking center of single pointing was set to ($\alpha_{\rm J2000.0}$, $\delta_{\rm J2000.0}$) 
= (23$^{\rm h}$03$^{\rm m}$15$^{\rm s}$.617, $+$08$\arcdeg$52$\arcmin$26$\arcsec$.06s), 
which was based on our previous ALMA observations at Band 7 \citep{2015ApJ...811...39I}. 
Each receiver was tuned to cover one of the above-mentioned lines in the 2 sideband dual-polarization mode. 
Each spectral window has a bandwidth of 1.875 GHz, and two windows 
were placed to each sideband (upper and lower) to achieve a total frequency coverage of $\sim 7.5$ GHz. 
We used two configurations of 12m arrays in Band 3 to 7 observations 
so that we can acquire both high angular resolutions sufficient 
to separate the starburst ring (radius $\sim 1\arcsec.5$) 
from the CND (radius $\sim 0\arcsec.5$) and recover 
the bulk of the emission extending over the central $\sim$a few arcsec scale. 
Note that, however, we decided not to use the compact configuration 
data of Band 7 observations as it would have issues in the calibration process. 
The Atacama Compact Array (ACA) was employed in the Band 8 observations for the same purpose: 
we did not include the total power array observations in this program. 
These result in the nominal maximum recoverable scales of our observations larger than 10$\arcsec$. 
Owing to this, as well as to the fact that we will focus on line ratios 
measured in several CO- or continuum-bright knots (i.e., compact structures), 
we do not consider effect of missing flux in this work. 
Further details of our observations are summarized in Table \ref{tbl3}. 

\begin{deluxetable}{ccccc}
\tabletypesize{\small}
\tablecaption{The Line Excitation Parameters\label{tbl2}}
\tablewidth{0pt}
\tablehead{
\colhead{\multirow{2}{*}{Line}} & \colhead{$\nu_{\rm rest}$} & \colhead{$E_u/k_{\rm B}$} & $A_{ul}$ & $n_{\rm cr}$ \\ 
 & (GHz) & (K) & (s$^{-1}$) & (cm$^{-3}$) 
}
\startdata
$^{12}$CO(1--0) & 115.2712 & 5.5 & 7.20 $\times$ 10$^{-8}$ & 2.1 $\times$ 10$^3$ \\
$^{13}$CO(2--1) & 220.3987 & 15.9 & 6.04 $\times$ 10$^{-7}$ & 9.7 $\times$ 10$^3$ \\
$^{12}$CO(2--1) & 230.5380 & 16.6 & 6.91 $\times$ 10$^{-7}$ & 1.1 $\times$ 10$^4$ \\
$^{12}$CO(3--2) & 345.7960 & 33.2 & 2.50 $\times$ 10$^{-6}$ & 3.6 $\times$ 10$^4$ \\
$[$\ion{C}{1}$]$(1--0) & 492.1607 & 23.6 & 7.88 $\times$ 10$^{-8}$ & 1.2 $\times$ 10$^3$ 
\enddata
\tablecomments{Values are adopted from the LAMDA database \citep{2005AA...432..369S}. 
$E_u$ and $k_{\rm B}$ are the upper energy level from the ground state and the Boltzmann constant. 
The critical densities are calculated for $T_{\rm kin} = 100$ K in the optically thin limit, 
simply as $n_{\rm cr} = A_{ul}/C_{ul}$, where $A_{ul}$ and $C_{ul}$ 
are the Einstein A- and C-coefficients of a $u \rightarrow l$ transition. 
We only considered H$_2$ as a collision partner.}
\end{deluxetable}

\begin{deluxetable*}{cccccccc}
\tabletypesize{\small}
\tablecaption{Journal of Our ALMA Observations\label{tbl3}}
\tablewidth{0pt}
\tablehead{
\colhead{\multirow{2}{*}{Line}} & \colhead{Date} & \colhead{Antenna} & \colhead{Baseline} & Integration & \multicolumn{3}{c}{Calibrator} \\ \cline{6-8}
 & \colhead{(UT)} & \colhead{Number} & \colhead{(m)} & \colhead{(min)} & Bandpass & Flux & Phase 
}
\decimalcolnumbers
\startdata
\multirow{2}{*}{$^{12}$CO(1--0) (Band 3)} & 2017 Dec 07 & 46 & 41--3600 & 38 & J2253$+$1608 & J2253$+$1608 & J2257$+$0743 \\
 & 2018 Apr 25 & 42 & 15--500 & 15 & J0006$-$0623 & J0006$-$0623 & J2257$+$0743 \\ \tableline 
\multirow{4}{*}{$^{13}$CO(2--1) (Band 6)} & 2018 Jun 03 & 47 & 15--361 & 38 & J2148$+$0657 & J2148$+$0657 & J2257$+$0743 \\
 & 2018 Jun 08 & 43 & 15--314 & 24 & J2253$+$1608 & J2253$+$1608 & J2257$+$0743 \\ 
 & 2018 Sep 15 & 44 & 15--1261 & 55 & J2253$+$1608 & J2253$+$1608 & J2257$+$0743 \\
 & 2018 Sep 15 & 44 & 15--1261 & 54 & J2253$+$1608 & J2253$+$1608 & J2257$+$0743 \\ \tableline 
\multirow{3}{*}{$^{12}$CO(2--1) (Band 6)} & 2018 May 22 & 46 & 15--314 & 45 & J2253$+$1608 & J2253$+$1608 & J2257$+$0743 \\ 
 & 2018 Sep 14 & 43 & 15--1230 & 63 & J2253$+$1608 & J2253$+$1608 & J2257$+$0743 \\
 & 2018 Sep 14 & 43 & 15--1230 & 77 & J2253$+$1608 & J2253$+$1608 & J2257$+$0743 \\ \tableline 
\multirow{2}{*}{$^{12}$CO(3--2) (Band 7)} & 2018 Jul 16$^\dag$ & 47 & 15--314 & 3.5 & J2253$+$1608 & J2253$+$1608 & J2232$+$1143 \\ 
 & 2018 Sep 05 & 46 & 15--784 & 28 & J2253$+$1608 & J2253$+$1608 & J2320$+$0513 \\ \tableline 
\multirow{10}{*}{[\ion{C}{1}](1--0) (Band 8)} & 2018 Aug 22 & 48 & 15--484 & 85 & J2253$+$1608 & J2253$+$1608 & J2232$+$1143 \\ 
 & 2018 Aug 22 & 48 & 15--484 & 85 & J2253$+$1608 & J2253$+$1608 & J2320$+$0513 \\ 
 & 2018 May 12 & 10 & 9--49 & 50 & J2258$-$2758 & J2258$-$2758 & J2253$+$1608 \\ 
 & 2018 May 13 & 11 & 9--49 & 50 & J2258$-$2758 & J2258$-$2758 & J2253$+$1608 \\  
 & 2018 May 19 & 12 & 9--49 & 50 & J2258$-$2758 & J2258$-$2758 & J2253$+$1608 \\
 & 2018 May 22 & 12 & 9--49 & 50 & J2258$-$2758 & J2258$-$2758 & J2253$+$1608 \\ 
 & 2018 May 23 & 12 & 9--49 & 50 & J2258$-$2758 & J2258$-$2758 & J2253$+$1608 \\   
 & 2018 May 23 & 11 & 9--49 & 50 & J2258$-$2758 & J2258$-$2758 & J2253$+$1608 \\         
 & 2018 May 25 & 11 & 9--49 & 50 & J2258$-$2758 & J2258$-$2758 & J2253$+$1608 \\   
 & 2018 May 27 & 12 & 9--49 & 50 & J2258$-$2758 & J2258$-$2758 & J2253$+$1608 \\  
\enddata
\tablecomments{(1) observed line and the corresponding ALMA Band. 
(2) observing date in UT. (3) number of antennas used in the observation. 
(4) baseline length in meter. The minimum and maximum lengths are shown. 
(5) net on-source integration time in minute. (6)-(8) calibrators used in the observation. 
$^\dag$This dataset was found to have serious issues in the calibration process. 
Given the short integration time that would not influence the overall properties of the $^{12}$CO(3--2), 
we excluded this from our subsequent analysis.}
\end{deluxetable*}

\begin{deluxetable*}{cccccc}
\tabletypesize{\small}
\tablecaption{Achieved Cube Parameters\label{tbl4}}
\tablewidth{0pt}
\tablehead{
\colhead{\multirow{2}{*}{Emission}} & \colhead{Beam} & \colhead{Beam} & rms &\colhead{$dV$} & Peak value \\ 
 & \colhead{($\arcsec \times \arcsec$) ($^\circ$)} & \colhead{(pc $\times$ pc)} & \colhead{(mJy beam$^{-1}$)} & (km s$^{-1}$) & (mJy beam$^{-1}$) 
}
\startdata
$^{12}$CO(1--0) & 0.36 $\times$ 0.29 ($-$72.5$\arcdeg$) & 120 $\times$ 96 & 0.46 & 10 & 23.8 \\ 
$^{13}$CO(2--1) & 0.35 $\times$ 0.28 (41.2$\arcdeg$) & 117 $\times$ 92 & 0.13 & 20 & 10.0 \\ 
$^{12}$CO(2--1) & 0.37 $\times$ 0.30 (51.2$\arcdeg$) & 123 $\times$ 101 & 0.28 & 10 & 96.0 \\ 
$^{12}$CO(3--2) & 0.37 $\times$ 0.31 ($-$87.9$\arcdeg$) & 122 $\times$ 102 & 1.31 & 10 & 201 \\ 
$[$\ion{C}{1}$]$(1--0) & 0.34 $\times$ 0.31 (75.7$\arcdeg$) & 113 $\times$ 101 & 3.04 & 10 & 146 \\ \tableline 
860 $\mu$m continuum & 0.34 $\times$ 0.28 ($-$87.3$\arcdeg$) & 113 $\times$ 93 & 0.07 & - & 5.16 \\ 
\enddata
\end{deluxetable*}

The reduction and calibration were done with CASA version 5.4 
\citep{2007ASPC..376..127M} in the standard manner by using the CASA pipeline. 
The continuum emission was identified and subtracted 
in the $uv$-plane for each visibility set of different array configuration, 
and the resultant visibilities were properly combined by the task \verb|concat|. 
All of the images presented in this paper were reconstructed 
using the task \verb|clean| with the Briggs weighting. 
We applied the \verb|robust| parameter of 0.0 to the Band 6, 7, and 8 datasets, 
which resulted in angular resolutions of $\sim 0\arcsec.34-0\arcsec.37$. 
As the same \verb|robust| parameter produced a much higher resolution for the Band 3 data, 
we instead adopted \verb|robust| = $+$0.5 with further tapering (\verb|outertaper| = 0$\arcsec$.2) 
to match its native beam size to the others as much as possible. 
The final achieved angular resolutions and 1$\sigma$ sensitivities are listed in Table \ref{tbl4}. 
Here the rms values were measured in channels free of line emission. 
Note that the channel spacings were originally 1.953 MHz ($^{12}$CO 
and $^{13}$CO lines) and 3.906 MHz ([\ion{C}{1}](1--0)), 
but we created cubes of $^{12}$CO and [\ion{C}{1}](1--0) lines 
with a common velocity resolution of $dV = 10$ km s$^{-1}$ to improve the signal-to-noise (S/N) ratios. 
For the case of the $^{13}$CO(2--1) cube, we adopted $dV = 20$ km s$^{-1}$ due to the faintness of the line. 
In the following, we express velocities in the optical convention with respect to the local standard of rest. 
These data products were further analyzed with the MIRIAD software \citep{1995ASPC...77..433S}. 

We also reconstructed five continuum maps by using 
each of the five datasets of different frequencies listed in Table \ref{tbl3}. 
The rms values of these maps were measured in areas free of emission. 
By inspecting the maps, we noticed that the notable peak positions 
(both in the CND and the starburst ring) of the Band 8 continuum map, 
as well as [\ion{C}{1}](1--0) map, are offset by $\sim 0\arcsec.08$ 
($\sim 20\%$ of the synthesized beam) toward the north-east direction 
relative to the corresponding positions in the other maps. 
This is due to a phase error in the Band 8 dataset, 
as confirmed by inspecting the position of a dedicated calibration source (quasar). 
For the analysis in the following, we corrected this positional offset. 
Note that in this work, we will use only the Band 7 (860 $\mu$m) continuum map 
to define representative locations to measure line fluxes. 
Details of the full continuum properties will be discussed elsewhere. 

Throughout the paper, we use line intensities corrected 
for the primary beam attenuation for quantitative discussion, 
but this correction is not critical as most of the emission 
is within $r \lesssim 2\arcsec$ from the center, which is well 
smaller than the primary beams (see also Figure \ref{fig1}). 
We simply refer to the $^{12}$CO isotopologue as CO, 
whereas $^{13}$CO is explicitly identified. 
The atomic carbon species is denoted as C$^0$, with its fine structure transition as [\ion{C}{1}]. 
The pixel scale of ALMA images is set to 0$\arcsec$.06, 
and the displayed errors indicate only statistical ones unless mentioned otherwise. 
As for the systematic error, the absolute flux calibration uncertainty is $\sim 10\%$ 
according to the Cycle 5 ALMA Proposer's Guide.

\section{Results}\label{sec3}
\subsection{Spatial distributions}\label{sec3.1}
\begin{figure*}
\begin{center}
\includegraphics[scale=0.55]{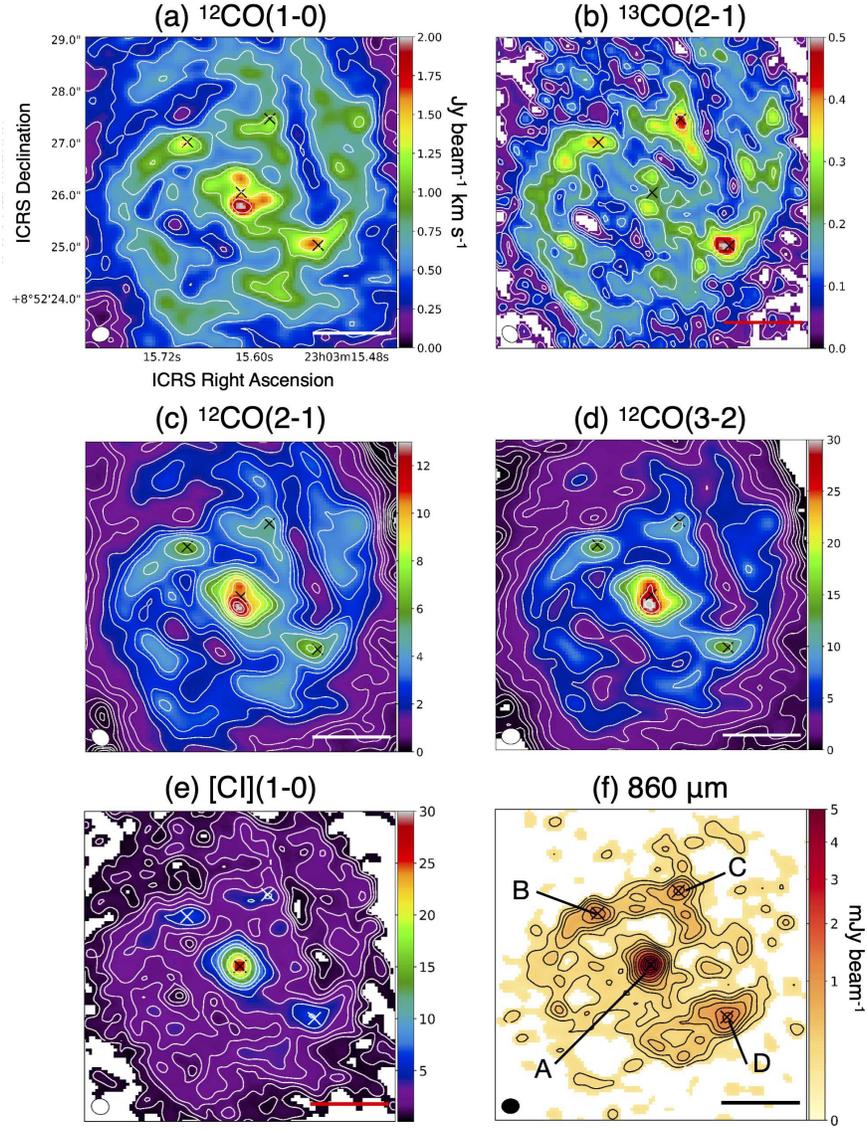}
\caption{
Integrated intensity maps of (a) $^{12}$CO(1--0), (b) $^{13}$CO(2--1), 
(c) $^{12}$CO(2--1), (d) $^{12}$CO(3--2), and (e) [\ion{C}{1}](1--0), 
in the central $\sim 2$ kpc region of NGC 7469, 
shown in the Jy beam$^{-1}$ km s$^{-1}$ unit. 
The common velocity range integrated over is 4650--5200 km s$^{-1}$ for all line emission maps. 
The 860 $\mu$m continuum emission map (mJy beam$^{-1}$ unit) is also shown 
in the panel (f) to define the four representative positions (A--D; shown by the crosses in each panel), 
where we measure line fluxes and spectra. 
The position-A corresponds to the AGN position of this galaxy. 
In the emission line maps, contours are drawn at 3, 5, 7, 10, 15, 20, 30, 40, 50, 75, 100, 
125, 150, 200, 250, 300, 350, ..., 600, and 650$\times \sigma$ levels 
(1$\sigma$ values are listed in Table \ref{tbl5}). 
The bottom-left filled ellipses indicate the synthesized beams 
and the horizontal bar in each panel corresponds to 500 pc length. 
Note that the native resolution maps are shown here, whereas we convolve 
these maps to a common 0.38$\arcsec$ resolution to take line ratios. 
The signals below 1.5$\sigma$ are masked to enhance the clarity. 
}
\label{fig2}
\end{center}
\end{figure*}

We first describe the distribution of cold ISM traced by CO, $^{13}$CO, and [\ion{C}{1}] lines. 
Figure \ref{fig2} shows the velocity-integrated intensity maps 
toward a central 6$\arcsec$ ($\sim 2$ kpc) boxy region of NGC 7469. 
We integrated a common velocity range of 4650--5200 km s$^{-1}$, 
which surely covers the line spectra at the nucleus (\S~3.2), 
without any clipping to make unbiased images, by using the MIRIAD task \verb|moment|. 
The zeroth moment is defined as $I = dV~\Sigma_i S_i$, where $S$ is the line intensity 
and the summation is taken over the $i$th velocity channel of width $dV$. 
Properties of these maps can be found in Table \ref{tbl5}. 
Emission of these lines were significantly detected 
both from the starburst ring (hereafter denoted as SB ring) 
and the CND (central $\sim 1\arcsec$), as well as the regions between or outside of them. 
Note that the SB ring is actually composed of two major spiral arms. 

Among the multiple transitions of CO, $J$ = 1--0 and 2--1 lines are frequently used 
as a tracer of bulk molecular gas due to their low $n_{\rm cr}$ \citep{2008AJ....136.2782L,2013AJ....146...19L}. 
These lines are brighter at the CND than at the SB ring, manifesting a large amount of cold molecular gas in the CND. 
Contrary to this base distribution, we notice that the $^{13}$CO(2--1) emission is significantly fainter at the CND than at the SB ring. 
Similar low intensities of $^{13}$CO relative to $^{12}$CO have been observed 
in SB systems with high FIR luminosities \citep[e.g.,][]{1991A&A...249..323A,1995A&A...300..369A,1998A&A...329..443H}. 
As discussed in \citet{2018PASJ...70L...1M} for the case of the low-luminosity AGN NGC 613, 
we consider this faintness of $^{13}$CO(2--1) in NGC 7469 is due to a low optical depth of the emission (see \S~4.1.2), 
and not due either to SB-induced selective photodissociation 
or nucleosynthesis \citep[e.g.,][]{1984ApJ...277..581L,1988ApJ...334..771V,1998ApJ...495..267M}: 
the latter two processes are inconsistent with the significantly lower star-formation activity 
in the CND than in the SB ring \citep{2014ApJ...780...86E}. 

In Figure \ref{fig2}f we also show the distribution of the 860 $\mu$m (Band 7) continuum emission, 
which defines four representative positions (A--D) to measure line fluxes and extract spectra. 
The coordinates of these positions are listed in Table \ref{tbl6}. 
The position-A coincides with the peak position of the Very Large Array 
8.4 GHz (3.5 cm) continuum emission \citep{1991ApJ...378...65C,2010MNRAS.401.2599O} within $\sim 0\arcsec.1$. 
We then regard this 860 $\mu$m peak position as the AGN position of NGC 7469. 
The 860 $\mu$m continuum distribution in the SB ring appears well consistent with those of 
MIR \citep{2003AJ....126..143S}, indicating that it traces thermal dust emission heated by young stars. 
Note however, we would need careful modelings 
of the continuum spectral energy distribution (SED) 
at the position-A to reveal its exact origin, as non-thermal synchrotron emission 
can be a severe contaminant even at Band 7, 
as observed in other galaxies \citep[e.g.,][]{2014A&A...567A.125G}. 
Such modeling will be presented elsewhere.

\begin{deluxetable}{ccc}
\tabletypesize{\small}
\tablecaption{Integrated Intensity Maps\label{tbl5}}
\tablewidth{0pt}
\tablehead{
\colhead{\multirow{2}{*}{Emission}} & \colhead{Beam} & \colhead{rms} \\ 
 & \colhead{($\arcsec \times \arcsec$)} & \colhead{(Jy beam$^{-1}$ km s$^{-1}$)}  
}
\startdata
\multirow{2}{*}{$^{12}$CO(1--0)} & 0.36 $\times$ 0.29 & 0.034 \\ 
 & 0.38 & 0.036 \\ \tableline
\multirow{2}{*}{$^{13}$CO(2--1)} & 0.35 $\times$ 0.28 & 0.014 \\ 
 & 0.38 & 0.015 \\ \tableline
\multirow{2}{*}{$^{12}$CO(2--1)} & 0.37 $\times$ 0.30 & 0.021 \\ 
 & 0.38 & 0.025 \\ \tableline
\multirow{2}{*}{$^{12}$CO(3--2)} & 0.37 $\times$ 0.31 & 0.098 \\ 
 & 0.38 & 0.118 \\ \tableline
\multirow{2}{*}{$[$\ion{C}{1}$]$(1--0)} & 0.34 $\times$ 0.31 & 0.227 \\ 
 & 0.38 & 0.245 \\ 
\enddata
\tablecomments{The velocity range of 4650--5200 km s$^{-1}$ was integrated to make these maps. 
Both the native resolution data and the common 0$\arcsec$.38 resolution data are presented.}
\end{deluxetable}

\begin{deluxetable}{ccc}
\tabletypesize{\small}
\tablecaption{Coordinates of the Peak Positions\label{tbl6}}
\tablewidth{0pt}
\tablehead{
\colhead{\multirow{2}{*}{Position}} & \colhead{R.A.} & \colhead{Dec.} \\ 
 & (ICRS) & (ICRS) 
}
\startdata
A & 23$^{\rm h}$03$^{\rm m}$15$^{\rm s}$.617 & $+$08$\arcdeg$52$\arcmin$26$\arcsec$.00 \\
B & 23$^{\rm h}$03$^{\rm m}$15$^{\rm s}$.686 & $+$08$\arcdeg$52$\arcmin$27$\arcsec$.02 \\
C & 23$^{\rm h}$03$^{\rm m}$15$^{\rm s}$.581 & $+$08$\arcdeg$52$\arcmin$27$\arcsec$.45 \\
D & 23$^{\rm h}$03$^{\rm m}$15$^{\rm s}$.518 & $+$08$\arcdeg$52$\arcmin$25$\arcsec$.02 \\
\enddata
\end{deluxetable}

\begin{figure}
\begin{center}
\includegraphics[width=\linewidth]{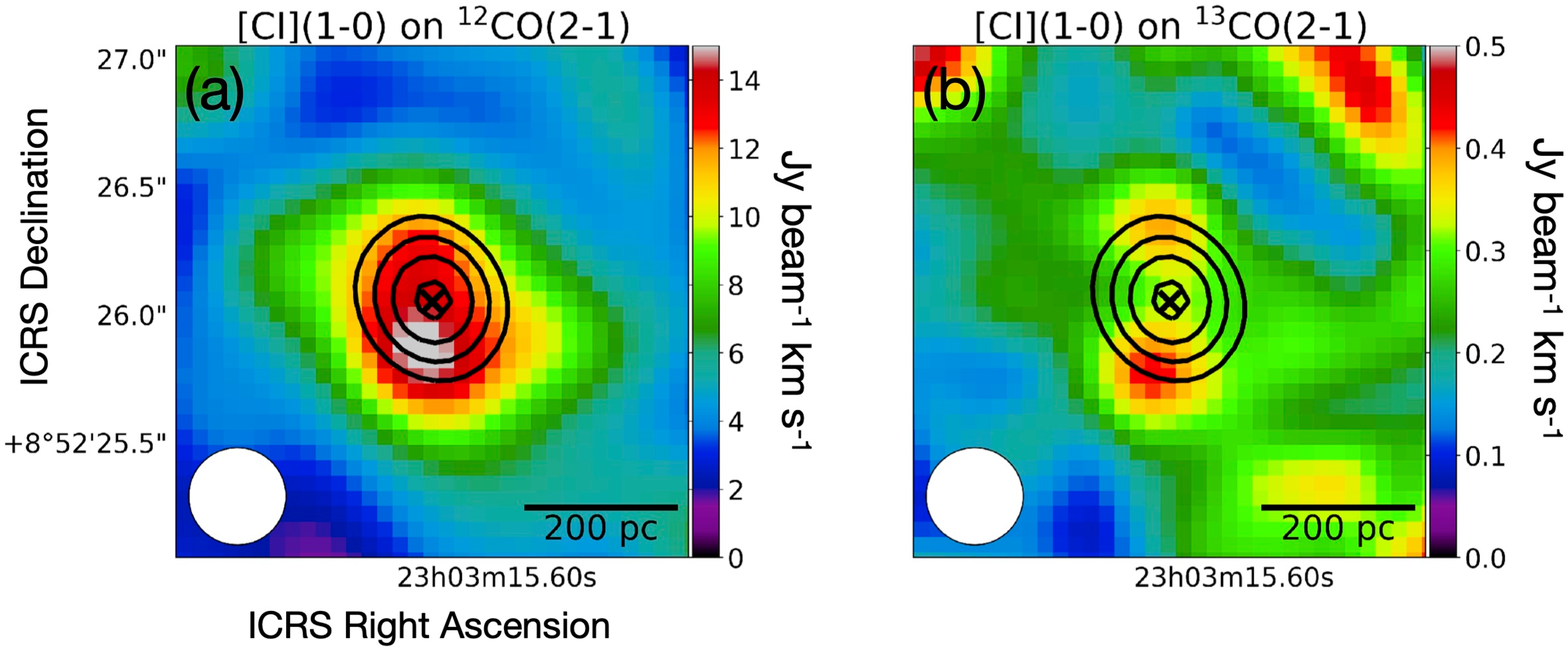}
\caption{
Closed-up view of the CND. 
The color scales indicate (a) CO(2--1) and (b) $^{13}$CO(2--1) distributions, respectively. 
The 1$\sigma$ rms values are 0.025 (a) and 0.015 (b) Jy beam$^{-1}$ km s$^{-1}$, respectively.  
Also plotted contours indicate [\ion{C}{1}](1--0) distribution 
(70, 80, 90, ..., 130$\sigma$, where 1$\sigma$ = 0.245 Jy beam$^{-1}$ km s$^{-1}$). 
These emissions are mapped at a common 0$\arcsec$.38 resolution. 
The central cross in each panel defines the AGN location (position-A). 
}
\label{fig3}
\end{center}
\end{figure}

In the SB ring all the line emission distributions peak at roughly the same positions 
as the 860 $\mu$m continuum emission, suggesting that these are star-forming giant molecular clouds (GMCs). 
On the other hand, there is a clear difference inside the CND between the CO lines 
(both $^{12}$CO and optically thinner $^{13}$CO) and the [\ion{C}{1}](1--0). 
The [\ion{C}{1}](1--0) distribution clearly peaks at the exact AGN position, 
whereas CO lines have two bright knots at $\sim 0\arcsec.2$ north and south to the AGN. 
This is not due to a slightly mismatched resolutions between the CO cubes and the [\ion{C}{1}](1--0) cube. 
Indeed, we still see the same spatial difference after convolving 
the resolutions to a common 0$\arcsec$.38 (Figure \ref{fig3}). 
Such a difference has not been the case in Galactic molecular clouds 
\citep[e.g.,][]{1997IAUS..178..129K,2000ApJ...539L.133P,2002ApJS..139..467I}, 
nor in nearby SB galaxies \citep{2016A&A...592L...3K,2019ApJ...887..143S}, 
where global [\ion{C}{1}](1--0) distribution resembles to those of low-$J$ CO lines. 
As this resemblance provides the backbone for [\ion{C}{1}](1--0) as a molecular mass tracer, 
our result may call into question its reliability near AGNs. 
Note that this relative faintness of the CO lines at the AGN positions is 
unlikely due to absorption effect, considering the type-1 Seyfert geometry where the AGN is directly visible. 

Moreover, as compared to the CO and $^{13}$CO lines, 
this [\ion{C}{1}](1--0) emission is more centrally-concentrated. 
For example, the relative fractions of the line fluxes measured 
at the central $r = 0\arcsec.5$ circular region to those measured at the central $r = 3\arcsec$ region are, 
7.2 $\pm$ 0.4\% for CO(1--0), 6.0 $\pm$ 0.2\% for $^{13}$CO(2--1), 
11.4 $\pm$ 0.1\% for CO(2--1), 14.6 $\pm$ 0.3\% for CO(3--2), 
but 20.0 $\pm$ 0.7\% for [\ion{C}{1}](1--0), respectively. 
The higher central concentration in CO(3--2) than CO(1--0) is 
a consequence of a higher gas excitation at the inner regions of galaxies. 
On the other hand, given the comparable $n_{\rm cr}$ (Table \ref{tbl2}) of CO(1--0) and [\ion{C}{1}](1--0), 
the significantly higher central concentration of the latter line stands out. 
We would need higher gas temperature (as the upper level energy is much higher for [\ion{C}{1}](1--0) than for CO(1--0); Table \ref{tbl2}) 
but also elevated C$^0$ abundance around the AGN to explain this peculiar behavior. 
Given these different spatial distributions and central concentrations, 
we argue that the AGN influences the [\ion{C}{1}](1--0) brightness 
likely in a form of XDR as discussed in \S~4 \citep{1996ApJ...466..561M,2005A&A...436..397M}.

\subsection{Spectra and Channel maps}\label{sec3.2}
\begin{figure}
\begin{center}
\includegraphics[scale=0.6]{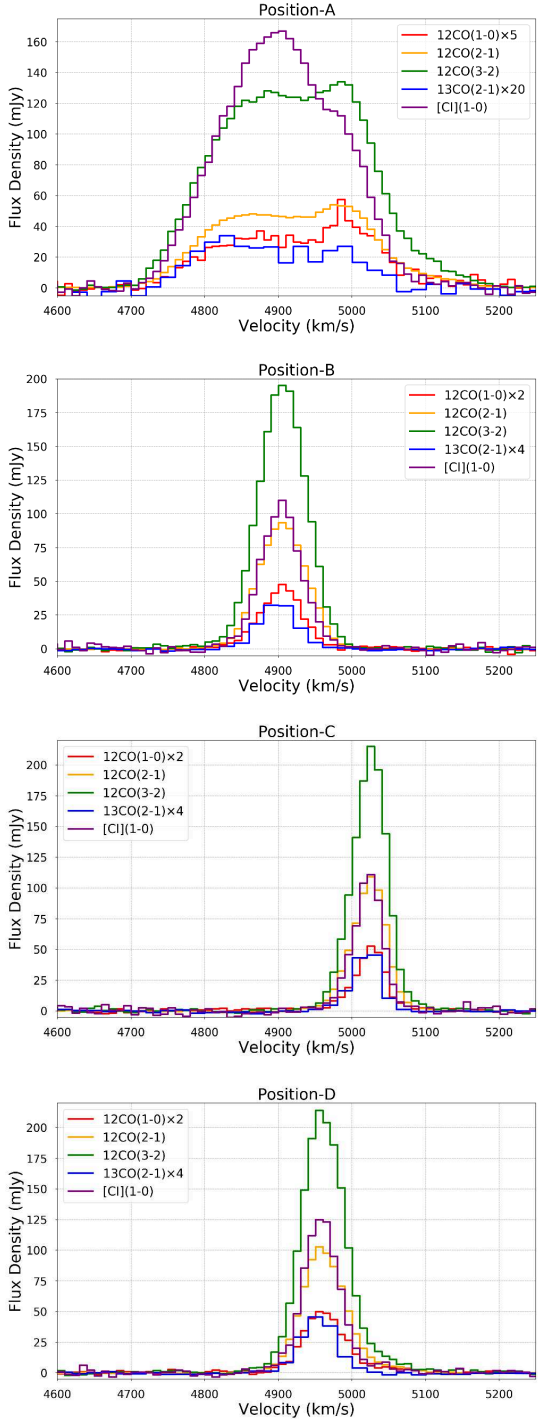}
\caption{
Continuum-subtracted spectra of $^{12}$CO(1--0), (2--1), (3--2), $^{13}$CO(2--1), and [\ion{C}{1}](1--0) 
extracted with the common 0$\arcsec$.38 ($\sim 130$ pc) aperture 
placed at the positions A--D (see Figure \ref{fig2}). 
Due to the faintness, $^{12}$CO(1--0) and $^{13}$CO(2--1) spectra 
are scaled by certain factors for a demonstrative purpose. 
}
\label{fig4}
\end{center}
\end{figure}

Figure \ref{fig4} compares the line spectra at the positions 
A--D measured with the common 0$\arcsec$.38 ($\sim 130$ pc) aperture. 
Note that the flux densities of CO(1--0) and $^{13}$CO(2--1) are multiplied 
by certain factors to fit into the panels due to their faintness. 
The lines are much broader at the position-A (full-width at zero-intensity FWZI $\sim 450$ km s$^{-1}$) 
than at the B--D (FWZI $\sim 150-200$ km s$^{-1}$). 
The different line widths indicate a higher turbulence 
at the position-A where the AGN reside than the rest positions, 
as well as likely higher enclosed mass within the aperture therein. 

\begin{figure*}
\begin{center}
\includegraphics[scale=0.5]{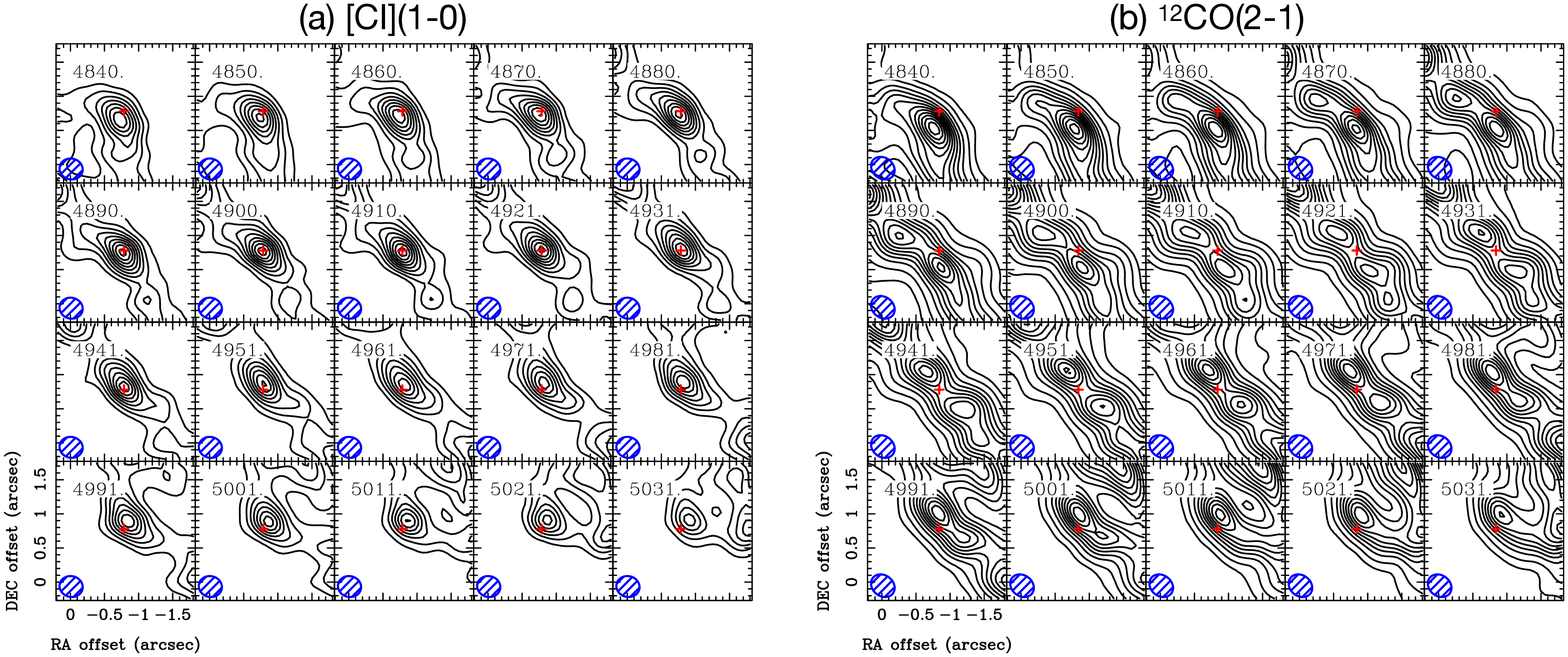}
\caption{
Velocity channel maps of (a) [\ion{C}{1}](1--0) and (b) $^{12}$CO(2--1) line emission 
in the central $2\arcsec \times 2\arcsec$ (1$\arcsec$ = 334 pc) region of NGC 7469. 
The red plus indicates the AGN position. 
The synthesized beam is plotted in the bottom-left corner. 
Contours are drawn at 5, 10, 15, .., 45$\sigma$ levels (1$\sigma$ = 3.04 mJy beam$^{-1}$) for (a), 
and 10, 25, 50, 75, ..., 300$\sigma$ levels (1$\sigma$ = 0.28 mJy beam$^{-1}$) for (b), respectively. 
Note that the positional offsets are relative to the phase reference center, 
which is not identical to the AGN position. 
At $V_{\rm LSR} \sim 4920$ km s$^{-1}$ the peak position 
of the [\ion{C}{1}](1--0) line distribution coincides with the exact AGN position.}
\label{fig5}
\end{center}
\end{figure*}

The CO and $^{13}$CO line profiles at the position-A clearly deviate from a single Gaussian, 
having two peaks at $V_{\rm LSR} \sim 4850$ km s$^{-1}$ and $V_{\rm LSR} \sim 4980$ km s$^{-1}$. 
The [\ion{C}{1}](1--0) profile also shows a deviation from a single Gaussian 
but with a less-prominent higher-velocity peak than the CO lines. 
The lower-velocity peak of the [\ion{C}{1}](1--0) line profile 
comes at around $V_{\rm LSR} \sim 4900$ km s$^{-1}$, which is offset to those of the CO lines. 
Such different line profiles between CO lines and [\ion{C}{1}](1--0) 
have not been clearly observed in, e.g., a SB galaxy NGC 1808 \citep{2019ApJ...887..143S} 
and the Large Magellanic Cloud \citep{2019A&A...621A..62O}. 
This difference is due to the different gas distributions in the velocity space 
as can be seen in the line channel maps (Figure \ref{fig5}; 
we present only the [\ion{C}{1}](1--0) and the CO(2--1) maps for simplicity). 

From the channel maps, it is evident that the [\ion{C}{1}](1--0) shows a rotating structure around the AGN, 
and peaks exactly at the AGN position at $V_{\rm LSR} \sim 4920$ km s$^{-1}$, 
whereas CO(2--1) does not show such a clear peak at the AGN. 
This velocity (4920 km s$^{-1}$) is roughly the average of the two peak velocities 
of the CO line profiles fitted with a double-Gaussian function (see Table \ref{tbl7}). 
It is also consistent with the previous estimate on the systemic velocity ($V_{\rm sys}$) of NGC 7469 \citep{1990ApJ...354..158M}, 
who defined $V_{\rm sys}$ as the line center of a CO(1--0) profile (4925 km s$^{-1}$). 
Given the higher resolution and the higher S/N ratio we obtained than previous sub/mm works, 
as well as our suggestion that the [\ion{C}{1}](1--0) brightness would reflect the AGN influence (\S~4.1), 
we decide to adopt the above $V_{\rm LSR} = 4920$ km s$^{-1}$ ($z = 0.01641$) 
as an updated $V_{\rm sys}$ throughout this work. 
This number is exactly the same as the $V_{\rm sys}$ that we dynamically estimate (\S~4.2).

We determine the line peak flux density, centroid velocity, 
full-width at half-maximum (FWHM), line flux, and line luminosity, 
by fitting a Gaussian function to the observed spectra. 
The results are summarized in Table \ref{tbl7}. 
Here we assumed a single Gaussian profile for the lines at the positions B--D, 
but we used a double Gaussian profile for those at the position-A, 
by taking the observed profiles into account. 
The line luminosity is calculated as 
\begin{equation}
\begin{split}
\left( \frac{L'_{\rm line}}{\rm K~km~s^{-1}~pc^2} \right) &= 3.25 \times 10^7 \left( \frac{D_L}{\rm Mpc} \right)^2 \left( \frac{\nu_{\rm rest}}{\rm GHz} \right)^{-2} \\ 
& \times (1+z)^{-1} \left( \frac{S\Delta V}{\rm Jy~km~s^{-1}} \right), 
\end{split} 
\end{equation}
where $S\Delta V$ is the line flux and $D_L$ is the luminosity distance \citep{2005ARA&A..43..677S}. 
The line luminosity is also computed in the unit of $L_\odot$ as 
\begin{equation}
\begin{split}
\left( \frac{L_{\rm line}}{L_\odot} \right) &= 1.04~\times~10^{-3} \left( \frac{\nu_{\rm rest}}{\rm GHz} \right) \\ 
& (1+z)^{-1} \left( \frac{S\Delta V}{\rm Jy~km~s^{-1}} \right) \left( \frac{D_L}{\rm Mpc} \right)^2
\end{split}
\end{equation}

In the SB ring, each line shows comparable fluxes 
among the positions B--D, implying similar ISM conditions therein. 
In terms of the line luminosity ($L_\odot$ unit), CO(3--2) clearly overwhelms the others, and [\ion{C}{1}](1--0) follows. 
The FWHM of the [\ion{C}{1}](1--0) is close to those of the CO(1--0) and $^{13}$CO(2--1), 
i.e., low excitation line or optically thin line. 
These FWHMs are smaller than those of the CO(2--1) and (3--2) lines. 
As the gas density is usually high ($\gtrsim 10^4$ cm$^{-3}$) in the central 
kpc regions of galaxies \citep[e.g.,][]{2014A&A...570A..28V}, 
one potential reason for these different line FWHMs is an opacity broadening (saturation effect) 
as bulk of the CO molecules can be excited to higher-$J$ levels. 

In the CND (position-A), there are clearly CO-weak and C$^0$-prominent 
velocity channels at around $V_{\rm LSR} \sim 4900$ km s$^{-1}$. 
Hence a special care would be required when we take line ratios at this AGN position. 
We found that both the lower-velocity component and the higher-velocity one 
have comparable line fluxes and FWHMs for the cases of the CO and $^{13}$CO lines. 
On the other hand, the lower-velocity component, which is closer to our $V_{\rm sys}$, 
is much brighter and wider for the case of the [\ion{C}{1}](1--0) line. 
It is also noteworthy that the [\ion{C}{1}](1--0) line luminosity ($L_\odot$ unit) is outstandingly high at this position: 
$L_{\rm line}$ of [\ion{C}{1}](1--0) is $(8.2 \pm 0.2) \times 10^4~L_\odot$ after summing up both the low and high velocity components, 
while it is $(7.6 \pm 0.1) \times 10^4~L_\odot$ even after adding all the CO and $^{13}$CO line luminosities observed here. 
Hence [\ion{C}{1}](1--0) contributes to the ISM cooling as significant as low-$J$ CO lines, 
implying that the chemical composition is different at this position-A as compared to the other positions in the SB-ring.

\begin{deluxetable*}{cc|cccc}
\tabletypesize{\small}
\tablecaption{Results of the Gaussian fitting\label{tbl7}}
\tablewidth{0pt}
\tablehead{
 & & \colhead{A} & \colhead{B} & \colhead{C} & \colhead{D}
}
\startdata
\multirow{6}{*}{$^{12}$CO(1--0)} & Peak (mJy) & 6.6 $\pm$ 0.2, 8.1 $\pm$ 0.3 & 22.6 $\pm$ 0.3 & 25.4 $\pm$ 0.3 & 24.9 $\pm$ 0.3 \\ 
& Line center (km s$^{-1}$) & 4862.0 $\pm$ 4.6, 5003.2 $\pm$ 2.8 & 4909.2 $\pm$ 0.4 & 5030.3 $\pm$ 0.3 & 4966.8 $\pm$ 0.3 \\
& FWHM (km s$^{-1}$) & 149.7 $\pm$ 9.5, 107.1 $\pm$ 5.3 & 67.4 $\pm$ 0.9 & 48.3 $\pm$ 0.7 & 67.5 $\pm$ 0.8 \\ 
& Flux (Jy km s$^{-1}$) & 1.0 $\pm$ 0.1, 0.9 $\pm$ 0.1 & 1.6 $\pm$ 0.1 & 1.3 $\pm$ 0.1 & 1.7 $\pm$ 0.1 \\ 
& $L'_{\rm line}$ (10$^6$ K km s$^{-1}$ pc$^2$) & 12.3 $\pm$ 0.9, 10.9 $\pm$ 0.7 & 19.0 $\pm$ 0.3 & 15.3 $\pm$ 0.3 & 20.9 $\pm$ 0.3 \\ 
& $L_{\rm line}$ (10$^2$ $L_\odot$) & 6.0 $\pm$ 0.4, 5.3 $\pm$ 0.4 & 9.3 $\pm$ 0.2 & 7.5 $\pm$ 0.1 & 10.2 $\pm$ 0.2 \\ \tableline 
\multirow{6}{*}{$^{13}$CO(2--1)} & Peak (mJy) & 1.6 $\pm$ 0.1, 1.2 $\pm$ 0.1 & 8.7 $\pm$ 0.1 & 12.8 $\pm$ 0.1 & 11.8 $\pm$ 0.1 \\ 
& Line center (km s$^{-1}$) & 4842.8 $\pm$ 5.1, 4981.9 $\pm$ 6.8 & 4909.2 $\pm$ 0.4 & 5030.0 $\pm$ 0.2 & 4964.2 $\pm$ 0.3 \\
& FWHM (km s$^{-1}$) & 113.5 $\pm$ 11.2, 113.1 $\pm$ 14.8 & 58.8 $\pm$ 0.9 & 44.2 $\pm$ 0.5 & 56.3 $\pm$ 0.6 \\ 
& Flux (Jy km s$^{-1}$) & 0.19 $\pm$ 0.02, 0.14 $\pm$ 0.02 & 0.52 $\pm$ 0.01 & 0.57 $\pm$ 0.01 & 0.67 $\pm$ 0.01 \\ 
& $L'_{\rm line}$ (10$^6$ K km s$^{-1}$ pc$^2$) & 0.63 $\pm$ 0.07, 0.47 $\pm$ 0.07 & 1.7 $\pm$ 0.1 & 1.9 $\pm$ 0.1 & 2.2 $\pm$ 0.1 \\ 
& $L_{\rm line}$ (10$^2$ $L_\odot$) & 2.2 $\pm$ 0.2, 1.6 $\pm$ 0.2 & 6.0 $\pm$ 0.1 & 6.6 $\pm$ 0.1 & 7.7 $\pm$ 0.1 \\ \tableline 
\multirow{6}{*}{$^{12}$CO(2--1)} & Peak (mJy) & 45.2 $\pm$ 0.2, 50.3 $\pm$ 0.2 & 93.7 $\pm$ 0.2 & 106.1 $\pm$ 0.2 & 101.5 $\pm$ 0.2 \\ 
& Line center (km s$^{-1}$) & 4847.7 $\pm$ 0.6, 4988.2 $\pm$ 0.6 & 4910.7 $\pm$ 0.1 & 5029.1 $\pm$ 0.1 & 4966.2 $\pm$ 0.1 \\
& FWHM (km s$^{-1}$) & 133.2 $\pm$ 1.0, 137.0 $\pm$ 0.9 & 76.3 $\pm$ 0.2 & 54.4 $\pm$ 0.1 & 69.1 $\pm$ 0.1 \\ 
& Flux (Jy km s$^{-1}$) & 6.1 $\pm$ 0.1, 7.0 $\pm$ 0.1 & 7.3 $\pm$ 0.1 & 5.9 $\pm$ 0.1 & 7.1 $\pm$ 0.1 \\ 
& $L'_{\rm line}$ (10$^6$ K km s$^{-1}$ pc$^2$) & 18.7 $\pm$ 0.2, 21.4 $\pm$ 0.2 & 22.2 $\pm$ 0.1 & 17.9 $\pm$ 0.1 & 21.8 $\pm$ 0.1 \\ 
& $L_{\rm line}$ (10$^2$ $L_\odot$) & 73.3 $\pm$ 0.6, 83.9 $\pm$ 0.7 & 87.1 $\pm$ 0.2 & 70.3 $\pm$ 0.2 & 85.4 $\pm$ 0.2 \\ \tableline  
\multirow{6}{*}{$^{12}$CO(3--2)} & Peak (mJy) & 114.0 $\pm$ 1.4, 118.4 $\pm$ 1.7 & 197.3 $\pm$ 0.8 & 210.6 $\pm$ 1.0 & 213.7 $\pm$ 0.9 \\ 
& Line center (km s$^{-1}$) & 4856.7 $\pm$ 1.7, 4990.3 $\pm$ 1.5 & 4910.9 $\pm$ 0.2 & 5029.9 $\pm$ 0.1 & 4966.0 $\pm$ 0.1 \\
& FWHM (km s$^{-1}$) & 145.3 $\pm$ 2.5, 133.7 $\pm$ 2.1 & 75.8 $\pm$ 0.4 & 54.3 $\pm$ 0.3 & 68.3 $\pm$ 0.3 \\ 
& Flux (Jy km s$^{-1}$) & 16.8 $\pm$ 0.4, 16.1 $\pm$ 0.3 & 15.2 $\pm$ 0.1 & 11.6 $\pm$ 0.1 & 14.8 $\pm$ 0.1 \\ 
& $L'_{\rm line}$ (10$^6$ K km s$^{-1}$ pc$^2$) & 22.9 $\pm$ 0.5, 21.9 $\pm$ 0.5 & 20.7 $\pm$ 0.1 & 15.8 $\pm$ 0.1 & 20.2 $\pm$ 0.1 \\ 
& $L_{\rm line}$ (10$^2$ $L_\odot$) & 302.9 $\pm$ 6.3, 289.5 $\pm$ 6.2 & 273.5 $\pm$ 1.7 & 209.1 $\pm$ 1.5 & 270.0 $\pm$ 1.6 \\ \tableline  
\multirow{6}{*}{[\ion{C}{1}](1--0)} & Peak (mJy) & 167.0 $\pm$ 1.2, 39.3 $\pm$ 3.3 & 105.3 $\pm$ 1.8 & 109.7 $\pm$ 2.1 & 126.6 $\pm$ 1.9 \\ 
& Line center (km s$^{-1}$) & 4898.8 $\pm$ 1.3, 5008.2 $\pm$ 1.7 & 4908.4 $\pm$ 0.5 & 5027.2 $\pm$ 0.4 & 4963.3 $\pm$ 0.4 \\
& FWHM (km s$^{-1}$) & 173.0 $\pm$ 2.7, 63.5 $\pm$ 5.6 & 65.1 $\pm$ 1.3 & 46.4 $\pm$ 1.0 & 57.8 $\pm$ 1.0 \\ 
& Flux (Jy km s$^{-1}$) & 29.4 $\pm$ 0.5, 2.5 $\pm$ 0.3 & 7.0 $\pm$ 0.2 & 5.2 $\pm$ 0.2 & 7.4 $\pm$ 0.2 \\ 
& $L'_{\rm line}$ (10$^6$ K km s$^{-1}$ pc$^2$) & 19.7 $\pm$ 0.3, 1.7 $\pm$ 0.2 & 4.7 $\pm$ 0.1 & 3.5 $\pm$ 0.1 & 5.0 $\pm$ 0.1 \\ 
& $L_{\rm line}$ (10$^2$ $L_\odot$) & 751.7 $\pm$ 12.9, 64.9 $\pm$ 7.9 & 178.3 $\pm$ 4.7 & 132.5 $\pm$ 4.0 & 190.5 $\pm$ 4.4 \\ \tableline  
\enddata
\tablecomments{
These values are measured at the four 860 $\mu$m continuum peak positions (A--D; Figure \ref{fig2}) 
with the common 0$\arcsec$.38 ($\sim 130$ pc) aperture. 
The systematic uncertainties are not included in the flux values. 
At the position-A, we performed double Gaussian fittings considering the observed line profiles (Figure \ref{fig4}). 
}
\end{deluxetable*}

\subsection{The [\ion{C}{1}](1--0) diagnostics}\label{sec3.3}
By using the results of the Gaussian fitting (Table \ref{tbl7}), we measure line flux ratios
\footnote{We express line ratios in the brightness temperature ($T_{\rm B}$) unit with the Rayleigh-Jeans approximation.} at the positions A--D. 
At the position-A, we use the combined flux of the low and high velocity components for simplicity. 
Hence the ratios at that position reflect an averaged property over the 0$\arcsec$.38 ($\sim 130$ pc) area. 
Selected channel map-based values will also be shown in the following. 

We here investigate [\ion{C}{1}](1--0)/CO(2--1) ($\equiv R_{\rm CI/CO}$) 
and [\ion{C}{1}](1--0)/$^{13}$CO(2--1) ($\equiv R_{\rm CI/13CO}$) $T_{\rm B}$ ratios 
based on our motivation to study XDR effects on the surrounding gas, including the dissociation of CO molecules. 
A dependence of $R_{\rm CI/13CO}$ on the environments (PDR vs XDR) 
has been discussed both in observational works \citep{2002A&A...383...82I,2015A&A...578A..95I} 
and in chemical models \citep{2007A&A...461..793M}. 
Both [\ion{C}{1}](1--0) and $^{13}$CO(2--1) lines are expected 
to be at least moderately optically thin under a wide range of physical conditions that would be valid for nearby galaxies. 
Hence their ratio is highly sensitive not only to excitation conditions but also to their abundances. 
CO(2--1) line has a $\sim 10\times$ higher $n_{\rm cr}$ than [\ion{C}{1}](1--0) in the optically thin limit, 
but its effective $n_{\rm cr}$ after accounting for photon trapping effects 
would be comparable to the $n_{\rm cr}$ of [\ion{C}{1}](1--0) \citep{2019ApJ...887..143S}. 
Hence $R_{\rm CI/CO}$ may also be sensitive to an abundance ratio, 
although the excitation and opacity effects should be carefully considered. 

These line ratios at the positions A--D are summarized in Table \ref{tbl8}. 
The line ratios are comparable at the positions B--D: 
we also found that their $R_{\rm CI/CO}$ are comparable to that found in the central region 
of the nearby SB galaxy NGC 253 \citep{2016A&A...592L...3K} after assuming $T_{\rm CO(1-0)} = T_{\rm CO(2-1)}$. 
On the other hand, both ratios are significantly higher at the position-A. 
The $R_{\rm CI/CO}$ and $R_{\rm CI/13CO}$ at the position-A are $\sim 2.5\times$ and $\sim 9\times$ 
higher than the values in the SB ring, respectively. 
In addition to these, we measured channel-based line ratios at the position-A, 
for example at the channel of 4900 km s$^{-1}$ that shows the brightest 
[\ion{C}{1}](1--0) emission (Figure \ref{fig4}), to better reflect the different line profiles we observed. 
Now the ratios become even higher: the $R_{\rm CI/CO}$ and $R_{\rm CI/13CO}$ at the position-A 
are $\sim 4\times$ and $\sim 11\times$ higher than the SB ring values, respectively. 
Therefore, it is evident that the [\ion{C}{1}](1--0) flux is dramatically 
enhanced relative to the CO and $^{13}$CO fluxes around the AGN 
as compared to the cases in the SB ring. 

We have also listed single dish (SD)-based flux ratios of NGC 7469 
in Table \ref{tbl8}, which are taken from \citet{2015A&A...578A..95I}. 
These values are obtained by the {\it Herschel} satellite and the ground-based James Clerk Maxwell Telescope 
after matching the resolutions to the {\it Herschel} data ($\sim 35\arcsec$). 
However, as the bright sources of the molecular line emission are the CND and the SB ring \citep{2004ApJ...602..148D}, 
the SD-based ratios basically reflect the averaged ISM properties of these structures. 
Indeed, the SD values are intermediate between the ALMA-based values at the position-A and B--D. 
It is therefore worth emphasizing that the $R_{\rm CI/13CO}$ of NGC 7469 (AGN) is 
$>5\times$ higher than the SD-based ratio. 
This manifests the power and the necessity of the high angular resolutions provided by ALMA 
to spatially separate the regions with different heating sources, 
in particular a compact AGN-influenced region from extended SB regions, 
to measure line ratios that reflect the environment properly.

\begin{deluxetable}{ccc}
\tabletypesize{\small}
\tablecaption{[\ion{C}{1}](1--0)-related line ratios ($T_{\rm B}$ unit)\label{tbl8}}
\tablewidth{0pt}
\tablehead{
\colhead{Position} & \colhead{[\ion{C}{1}](1--0)/CO(2--1)} & \colhead{[\ion{C}{1}](1--0)/$^{13}$CO(2--1)}
}
\startdata
A & 0.53 $\pm$ 0.08 & 19.5 $\pm$ 3.3 \\ 
B & 0.21 $\pm$ 0.03 & 2.68 $\pm$ 0.39 \\
C & 0.19 $\pm$ 0.03 & 1.81 $\pm$ 0.26 \\
D & 0.23 $\pm$ 0.03 & 2.22 $\pm$ 0.32 \\ \hline 
A (ch)$^\dag$ & 0.78 $\pm$ 0.11 & 25.1 $\pm$ 4.4 \\ 
SD$^\ddag$ & 0.30 $\pm$ 0.06 & 3.63 $\pm$ 0.77 \\
\enddata
\tablecomments{All ratios are taken with the common 0$\arcsec$.38 aperture and include the systematic flux uncertainties. 
$^\dag$Channel map-based line ratios at the position-A. 
We measured these ratios at $V_{\rm LSR} = 4900$ km s$^{-1}$, 
i.e., at the channel where the [\ion{C}{1}](1--0) becomes brightest. 
$^\ddag$Ratios measured with single dish (SD) observations (Israel et al. 2015).} 
\end{deluxetable}

To compare the observed line ratios in NGC 7469 with those of other galaxies with various nuclear activities, 
we again compiled line flux data of [\ion{C}{1}](1--0), CO(2--1), and $^{13}$CO(2--1) from \citet{2015A&A...578A..95I}. 
The literature data was taken with ground-based single dish telescopes with apertures of $>22\arcsec$ (see their Table 5), 
hence basically probes spatial scales of $>$ several kpc. 
Their sample includes AGNs (NGC 1068, NGC 3079, NGC 4736, NGC 4945, M51, and the Circinus galaxy), 
SB galaxies (IC 10, NGC 253, NGC 660, IC 342, Henize 2-10, NGC 3628, NGC 4038, M83, and NGC 6946; 
these include low-ionization nuclear emission-line region (LINER) type galaxies as well), 
and quiescent galaxies (NGC 278, NGC 891, Maffei 2). 
Our classification of the nuclear type is based on the record in the NED database except for IC 10 and Maffei 2; 
we classified these as SB given their high nuclear star-formation rates \citep{1998ARA&A..36..435M,2008ApJ...675..281M}. 

\begin{figure}
\begin{center}
\includegraphics[width=\linewidth]{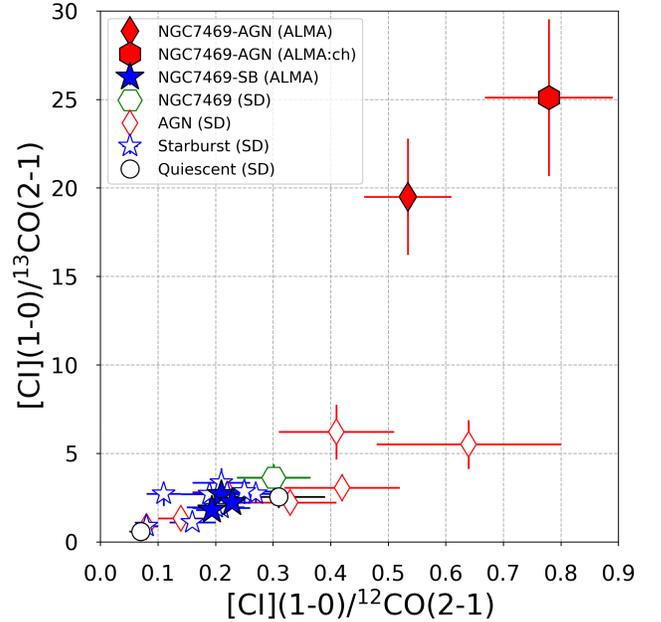}
\caption{
[\ion{C}{1}](1--0)/$^{12}$CO(2--1) = $R_{\rm CI/CO}$ 
vs [\ion{C}{1}](1--0)/$^{13}$CO(2--1) = $R_{\rm CI/13CO}$ line flux ratios ($T_{\rm B}$ scale). 
Our ALMA results of NGC 7469 are shown with the filled symbols, 
while single dish (SD) measurements are shown with the open symbols. 
The red diamonds, blue stars, black circles respectively indicate the data 
of AGN, SB galaxies, and quiescent galaxies. 
We also show the channel map-based ratios of the NGC 7469 AGN by the filled red hexagon. 
The SD measurements of NGC 7469 is also shown by the open green hexagon. 
Galaxies other than NGC 7469 (ALMA) are compiled from \citet{2015A&A...578A..95I}, 
all of which are based on SD observations. 
AGNs tend to have higher values in both ratios, 
but even among them the extremely high values of NGC 7469 (AGN) and NGC 7469 (AGN: ch) stand out. 
}
\label{fig6}
\end{center}
\end{figure}

The resultant plot of $R_{\rm CI/CO}$ vs $R_{\rm CI/13CO}$ is displayed in Figure \ref{fig6}. 
At first inspection, although the physical scales probed are different, 
one may see that some galaxies with AGN contribution tend to have higher ratios in both axes. 
The SB galaxies, as well as the SB ring of NGC 7469, are all clustered around 
$R_{\rm CI/CO} \sim 0.2-0.3$ and $R_{\rm CI/13CO} \sim 2-4$: 
physical and/or chemical conditions governing these regions/environments 
(e.g., PDR characteristics) are thus not likely different dramatically. 
While some SD-based AGN ratios are already significantly higher than those of the SB galaxies, 
our ALMA-based ratios of NGC 7469 AGN, particularly the channel map-based values, 
are outstandingly high in both $R_{\rm CI/CO}$ and $R_{\rm CI/13CO}$. 

In summary, to our best knowledge, these high ratios of NGC 7469 (AGN), 
or called as {\it \ion{C}{1}-enhancement} hereafter, have never been observed 
in SB galaxies or quiescent galaxies at the spatial scales probed here. 
Thus, we now consider that this diagram has a potential 
to discriminate nuclear activities, as a submm energy diagnostic tool.

\section{Discussion}\label{sec4}
In this section we investigate a physical origin of the \ion{C}{1}-enhancement revealed in \S~\ref{sec3.3} 
by performing both local thermodynamic equilibrium (LTE) and non-LTE analyses of $R_{\rm CI/CO}$ and $R_{\rm CI/13CO}$. 
The purpose of these analyses is to understand a trend of the underlying 
physical and/or chemical conditions to explain the \ion{C}{1}-enhancement. 
Further detailed non-LTE modeling with extensive comparison with chemical models 
will be presented in our forthcoming paper. 

One will later see in this section that an elevated C$^0$ abundance is required 
to explain the \ion{C}{1}-enhancement at the AGN position of NGC 7469. 
This calls a tension in molecular mass measurements using this line, 
particularly those at the CND-scale of AGN-host galaxies. 
Therefore, we dedicate \S~\ref{sec4.2} to derive a specific [\ion{C}{1}](1--0) 
to $M_{\rm H2}$ (and CO(1--0) to $M_{\rm H2}$) conversion factor based on our dynamical modeling.

\subsection{Physical origin of the CI-enhancement}\label{sec4.1}
\subsubsection{LTE perspective}\label{sec4.1.1}
We begin by calculating $R_{\rm CI/13CO}$ under optically thin LTE conditions 
to relate the flux ratio to C$^0$/CO column density ratios ($N_{\rm C^0}/N_{\rm CO}$). 
As these lines are likely optically thin or moderately opaque at the most physical conditions, 
as well as their $n_{\rm cr}$ are modest, 
we consider that both optically thin and LTE conditions are good approximations for a first-order estimation. 
From equations (1) and (2) of \citet{1995A&A...297..567T}, we can describe the line flux ratio as 
\begin{equation}\label{eq3}
R_{\rm CI/13CO} = 0.006~A^{12}_{13}~f(T_{\rm ex}) \times N_{\rm C^0}/N_{\rm CO}, 
\end{equation}
where 
\begin{equation}\label{eq4}
f(T_{\rm ex}) = T_{\rm ex}/(e^{7.0/T_{\rm ex}} + 3e^{-16.6/T_{\rm ex}} + 5e^{-55.5/T_{\rm ex}})
\end{equation}
with $T_{\rm ex}$ and $A^{12}_{13}$ denoting an excitation temperature (assumed to be common for all species) 
and an isotopic abundance ratio of [CO]/[$^{13}$CO], respectively. 
Hereafter [X] means an abundance of the species X. 
The isotopic ratio varies significantly from galaxy to galaxy, and even inside a single galaxy \citep[e.g.,][]{2005ApJ...634.1126M}. 
For example, $A^{12}_{13}$ (we assume that this is identical to [$^{12}{\rm C}]/[^{13}{\rm C}$] here) 
is $\sim 50-60$ at inner Galactic sources \citep{1998A&A...337..246L}, 
$\sim 25$ in the Galactic central region \citep{1985A&A...149..195G}, 
and $>40$ in nearby SB galaxies \citep{2010A&A...522A..62M,2014A&A...565A...3H,2019A&A...629A...6T}
\footnote{But see also \citet{2019A&A...624A.125M} for a smaller value of $A^{12}_{13} \sim 21$ observed in the SB galaxy NGC 253.}. 
Recently, \citet{2019A&A...629A...6T} measured this ratio in the type-2 Seyfert galaxy NGC 1068, 
which has a similar AGN luminosity to NGC 7469. 
As it is impractical to determine the isotopic ratio in NGC 7469 with the current dataset, 
we assume $A^{12}_{13} = 40$ hereafter, which is roughly the same value found in NGC 1068 ($\sim 38$). 
Note that, based on the equation (\ref{eq3}), $R_{\rm CI/13CO}$ linearly depends on the assumed $A^{12}_{13}$.

\begin{figure}
\begin{center}
\includegraphics[width=\linewidth]{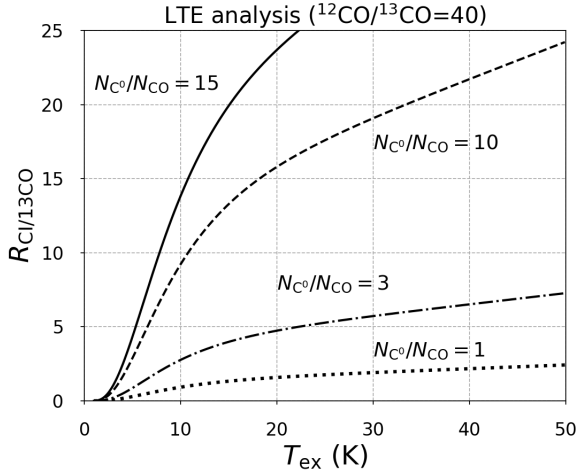}
\caption{
Expected [\ion{C}{1}](1--0)/$^{13}$CO(2--1) brightness temperature ratios 
as a function of line excitation temperature under the optically thin LTE condition. 
The four different curves respectively correspond to the case 
of $N_{\rm C^0}/N_{\rm CO} = 15$ (solid), = 10 (dashed), = 3 (dot-dashed), and = 1 (dotted). 
}
\label{fig7}
\end{center}
\end{figure}

By using the equations (\ref{eq3}) and (\ref{eq4}), we calculate 
the $R_{\rm CI/13CO}$ as a function of $T_{\rm ex}$, 
for varying $N_{\rm C^0}/N_{\rm CO}$, in Figure \ref{fig7}. 
Under these conditions, it is evident that the $R_{\rm CI/13CO}$ 
observed in the SB ring of NGC 7469 and other SB galaxies 
can be explained by $N_{\rm C^0}/N_{\rm CO} \lesssim 3$, 
but we would need further enhanced values, e.g., $N_{\rm C^0}/N_{\rm CO} \sim$ several to $\sim 15$ 
to explain the very high $R_{\rm CI/13CO}$ observed at the AGN position of NGC 7469. 
However, we emphasize that the actual ratio strongly depends on the $T_{\rm ex}$, 
which is hard to constrain by using the single transition [\ion{C}{1}](1--0) line and $^{13}$CO(2--1) line in our hand. 

\begin{figure}
\begin{center}
\includegraphics[width=\linewidth]{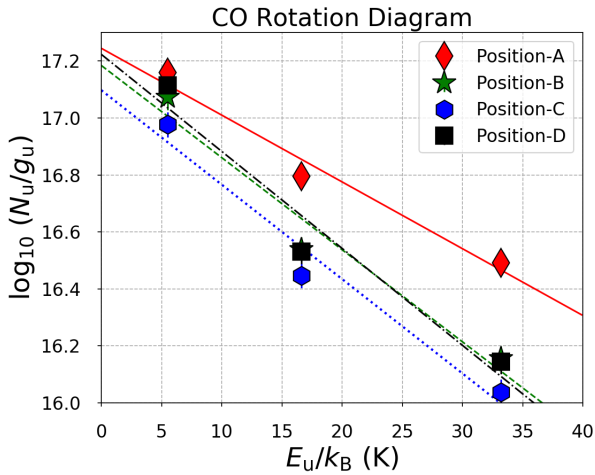}
\caption{
CO rotation diagrams of NGC 7469. 
The four different symbols and lines correspond to the cases at the positions A--D. 
The $T_{\rm ex}$ of each position derived from our linear regression fit are, 
18.5 $\pm$ 3.2 K (A), 13.4 $\pm$ 2.9 K (B), 13.1 $\pm$ 2.5 K (C), and 12.8 $\pm$ 3.0 K (D), respectively. 
}
\label{fig8}
\end{center}
\end{figure}

On the other hand, we may roughly estimate the $T_{\rm ex}$ at the positions A--D 
by constructing rotation diagrams \citep{1999ApJ...517..209G}. 
A rotation diagram is a plot of the column density per statistical weight of a number of molecular energy levels, 
as a function of their energies above the ground state. 
From the optically thin condition, the column density of the level $u$ ($N_u$) is written as 
\begin{equation}\label{eq5}
N_u = \frac{8 \pi k_{\rm B} \nu^2 \int T_{\rm B} dV}{hc^3 A_{ul}}, 
\end{equation}
where $k_{\rm B}$ and $h$ are the Boltzmann and Planck constants, $c$ is the speed of the light. 
Also, from the LTE condition, $N_u$ can be expressed as 
\begin{equation}\label{eq6}
N_u = \frac{N_{\rm X}}{Q(T_{\rm ex})} g_u \exp \left(  - \frac{E_u}{k_{\rm B} T_{\rm ex}}\right), 
\end{equation}
where $N_{\rm X}$ is the total column density of the given species X, 
$Q(T_{\rm ex})$ is a partition function ($= \Sigma_u g_u \exp(-E_u/k_{\rm B}T_{\rm ex})$), 
and $E_u$ is the energy at level $u$ from the ground state. 
Then the logarithm of $N_u/g_u$ vs $E_u/k_{\rm B}$ yields a straight line 
with a slope and the $y$-axis intercept indicative of $T_{\rm ex}$ 
(or rotation temperature $T_{\rm rot}$) and $N_{\rm X}$, respectively. 

By using the $J$ = 1--0, 2--1, and 3--2 CO line fluxes at the positions A--D, 
we constructed the rotation diagrams as shown in Figure \ref{fig8}. 
If we fit the data points by straight lines, inferred $T_{\rm ex}$ are 
18.5 $\pm$ 3.2 K (A), 13.4 $\pm$ 2.9 K (B), 
13.1 $\pm$ 2.5 K (C), and 12.8 $\pm$ 3.0 K (D), respectively. 
The estimated $T_{\rm ex}$ is relatively higher at the position-A than B--D, 
suggesting the existence of denser and/or warmer gas at the nucleus than at the SB ring. 
If these CO-based $T_{\rm ex}$ also hold for C$^0$ and $^{13}$CO excitations, 
our prediction on $N_{\rm C^0}/N_{\rm CO}$ ratios discussed above is valid. 
This would not be a very inappropriate speculation given the similar excitation conditions of the lines considered here (Table \ref{tbl2}). 
Thus, $N_{\rm C^0}/N_{\rm CO}$ is likely enhanced around the AGN as compared to the values at the SB ring. 

Note that, however, the rotation temperatures derived here 
ought to be considered as {\it lower limits} of the true excitation and kinetic temperatures due to finite optical depths. 
Our rotation diagrams are already suggestive of this as they are obviously curved. 
The curved feature implies either (i) some of the lines are at least moderately optically thick 
and (ii) there are multiple gas components with different excitation conditions. 
In the case of optically thick emission (case-(i)), the $N_{\rm u}/g_u$ value of a certain transition is underestimated 
by a factor of $C_\tau \equiv \tau/(1-e^{-\tau})$ \citep{1999ApJ...517..209G}: 
the actual $T_{\rm ex}$ critically depends on this $C_\tau$. 
In order at least to cope with this optical depth effect, 
we need to perform non-LTE analyses.

\subsubsection{Non-LTE perspective}\label{sec4.1.2}
Our non-LTE radiative transfer modelings are performed with the RADEX code \citep{2007A&A...468..627V} 
for spherical geometry to understand the underlying physical conditions of the \ion{C}{1}-enhancement, 
i.e., kinetic temperature ($T_{\rm kin}$), H$_2$ volume density ($n_{\rm H2}$), and $N_{\rm X}$. 
RADEX uses an escape probability approximation to treat optical depth effects and solves 
statistical equilibrium in a homogeneous (i.e., single temperature and density), one-phase medium. 
Thus, we need to assume that all lines observed are emitted from the same volume, 
although as we revealed in \S~\ref{sec3} the real structures are quite complex. 
Note that we do not intend to precisely model those complex environments here. 
The models described below are constructed for educated guesses of the relevant parameters 
\citep[see similar experiments in][]{2016ApJ...818...42I}. 

In our simulation, we investigated how the following parameters affect the line ratios of our interest. 
\begin{itemize}
\item[1.] Kinetic temperature ($T_{\rm kin}$): 
this affects the rate of the collisional excitation. 
The cases of 50, 100, 200, 300, and 500 K are investigated. 
This range mostly covers the CND-scale $T_{\rm kin}$ suggested for nearby AGNs and SB galaxies 
\citep[e.g.,][]{2008ApJ...677..262K,2013PASJ...65..100I,2014A&A...570A..28V}. 
\item[2.] Gas volume density ($n_{\rm H2}$): this also determines the rate of collisional excitation. 
Three cases of 10$^3$, 10$^4$, and 10$^5$ cm$^{-3}$ are studied. 
These are also typical values in the CNDs of nearby galaxies 
\citep[e.g.,][]{2008ApJ...677..262K,2013PASJ...65..100I,2014A&A...570A..28V}, 
as well as the values that can cover the $n_{\rm cr}$ of our target lines (Table \ref{tbl2}). 
\item[3.] Abundance ratio: throughout the work we assume [CO]/[$^{13}$CO] = 40 (\S~4.1.1). 
We studied three cases of [C$^0$]/[CO] = 1.0, 3.0, and 10.0. 
Note that [C$^0$]/[CO] $>1$ is required to reproduce $R_{\rm CI/13CO} >$ a few according to \citet{2002A&A...383...82I}. 
\item[4.] Optical depth ($\tau$): models with different $N_{\rm X}/\Delta V$ are used to test this effect. 
Here $\Delta V$ is the line velocity width, hence the ratio $N_{\rm X}/\Delta V$ is equivalently a ratio of 
a volume density of the target species to a velocity gradient over the line-of-sight. 
We set CO as our base species to consider this effect. 
For $N_{\rm CO}$ we made initial guesses from the observed CO fluxes. 
Applying the CO-to-H$_2$ conversion factor computed for the CND-scale of NGC 7469 \citep{2004ApJ...602..148D} 
to the CO(1--0) fluxes in Table \ref{tbl7}, we obtain $N_{\rm H2} = 1.4 \times 10^{23}$ cm$^{-2}$ at the position-A, 
as well as $N_{\rm H2} = (0.9-1.3) \times 10^{23}$ cm$^{-2}$ at the positions B--D, respectively. 
These translate into $N_{\rm CO} \sim 1 \times 10^{19}$ cm$^{-2}$ if we assume a typical [CO]/[H$_2$] abundance ratio of 10$^{-4}$. 
Considering this $N_{\rm CO}$ and the observed line widths, 
we here experimentally studied three cases of $N_{\rm CO}/\Delta V = 3 \times 10^{16}$ cm$^{-2}$ (km s$^{-1}$)$^{-1}$, 
$1 \times 10^{17}$ cm$^{-2}$ (km s$^{-1}$)$^{-1}$, and $3 \times 10^{17}$ cm$^{-2}$ (km s$^{-1}$)$^{-1}$. 
\item[5.] Background temperature ($T_{\rm bg}$): this affects radiative excitation rates of the lines. 
While we would expect high $T_{\rm bg}$ particularly around an AGN, 
we fix this to the cosmic microwave background temperature of 2.73 K in this work for simplicity. 
Note that however, this parameter potentially affects the resultant line excitation significantly, 
as sometimes radiative excitation becomes more important than collisional excitation \citep{2015ApJ...799...26M,2016ApJ...818...42I}. 
Indeed, as the upper level energy of [\ion{C}{1}](1--0) is higher than those of CO(2--1) and $^{13}$CO(2--1), 
both $R_{\rm CI/CO}$ and $R_{\rm CI/13CO}$ would become higher when we increase $T_{\rm bg}$. 
\end{itemize}

\begin{figure*}
\begin{center}
\includegraphics[width=\linewidth]{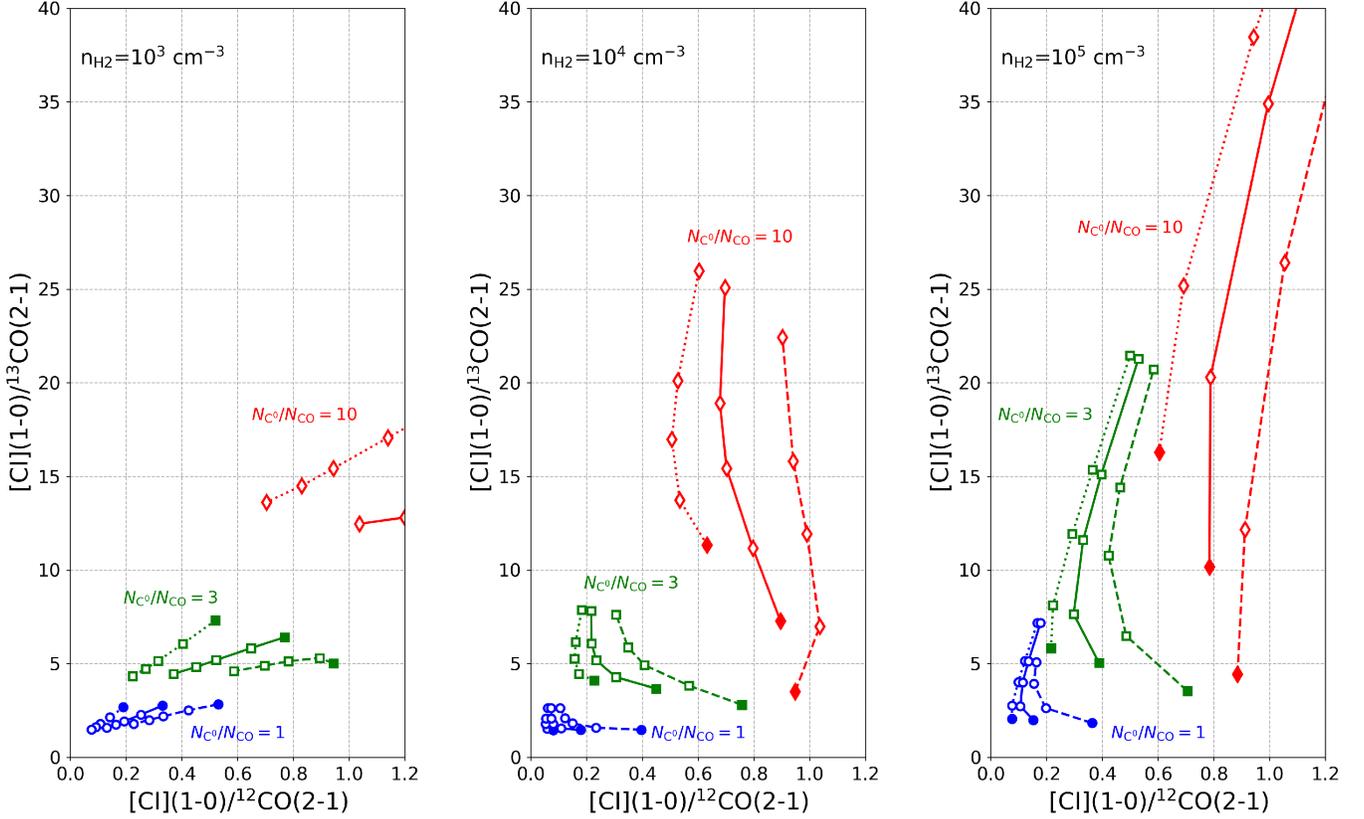}
\caption{
Radiative transfer modelings of [\ion{C}{1}](1--0)/CO(2--1) and [\ion{C}{1}](1--0)/$^{13}$CO(2--1) ratios ($T_{\rm B}$ unit). 
Three panels indicate the cases of different $n_{\rm H2}$. 
All model calculations assume the constant [$^{12}$CO]/[$^{13}$CO] = 40. 
In each panel, there are three groups of $N_{\rm C^0}/N_{\rm CO}$ (1, 3, and 10 shown in blue, green, and red colors, respectively), 
where different line tracks indicate different $N_{\rm CO}/\Delta V$: 
dotted = $3 \times 10^{16}$ cm$^{-2}$ (km s$^{-1}$)$^{-1}$, 
solid = $1 \times 10^{17}$ cm$^{-2}$ (km s$^{-1}$)$^{-1}$, 
and dashed = $3 \times 10^{17}$ cm$^{-2}$ (km s$^{-1}$)$^{-1}$, respectively. 
In each track, points indicate the five cases of $T_{\rm kin}$ (50, 100, 200, 300, and 500 K), 
with monotonically decreasing the [\ion{C}{1}](1--0)/$^{13}$CO(2--1) ratio with higher $T_{\rm kin}$ in the case of $n_{\rm H2} = 10^3$ cm$^{-3}$, 
or conversely, monotonically increasing the [\ion{C}{1}](1--0)/$^{13}$CO(2--1) ratio with higher $T_{\rm kin}$ in the remaining cases, respectively: 
we mark the lowest $T_{\rm kin} = 50$ K cases with filled symbols. 
}
\label{fig9}
\end{center}
\end{figure*}

The results of our radiative transfer calculations are summarized in Figure \ref{fig9}. 
It is evident that cases with higher $N_{\rm C^0}/N_{\rm CO}$ 
tend to show accordingly higher $R_{\rm CI/CO}$ and $R_{\rm CI/13CO}$. 
The $R_{\rm CI/13CO}$ is highly sensitive to both $T_{\rm kin}$ and $n_{\rm H2}$ in complex manners. 
In the high density cases of $n_{\rm H2} = 10^4$ and $10^5$ cm$^{-3}$, 
the $R_{\rm CI/13CO}$ monotonically increases in higher $T_{\rm kin}$. 
This is due to the fast reduction of $^{13}$CO(2--1) opacity and intensity 
as $^{13}$CO is easily excited to further upper rotational levels ($^{13}$CO(2--1) is always very optically thin in these cases). 
As for the [\ion{C}{1}](1--0), we found that its line intensity varies only slightly in each model track: 
although its line opacity reduces at some level, increasing $T_{\rm ex}$ eventually 
compensates the reduction to roughly maintain the resultant [\ion{C}{1}](1--0) intensity, 
which leads to the enhanced $R_{\rm CI/13CO}$ in higher excitation conditions. 
On the other hand, the reduction of the $^{13}$CO(2--1) opacity is 
only moderate in the cases of $n_{\rm H2} = 10^3$ cm$^{-3}$. 
Hence $^{13}$CO(2--1) now becomes brighter at higher $T_{\rm kin}$ (i.e., higher $T_{\rm ex}$), 
which resulted in the reduction of $R_{\rm CI/13CO}$ in these low density cases. 
Note that [\ion{C}{1}](1--0) is optically thin in most cases, but it can also be optically thick 
in limited situations, i.e., those under low excitation conditions (low $n_{\rm H2}$ and low $T_{\rm kin}$) 
with large $N_{\rm C^0}/\Delta V \gtrsim 10^{18}$ cm$^{-2}$ (km s$^{-1}$)$^{-1}$. 

The dependence of $R_{\rm CI/CO}$ on the excitation conditions 
is complex as our parameter space covers both optically thick and thin regimes of CO(2--1) emission: 
in the cases of $n_{\rm H2} = 10^3$ and $10^4$ cm$^{-3}$, 
$R_{\rm CI/CO}$ tends to decrease for higher $T_{\rm kin}$, 
as CO(2--1) intensity increases with $T_{\rm ex}$ in these high line opacity cases. 
Contrary to these, $R_{\rm CI/CO}$ turns to increase with $T_{\rm kin}$ 
when $n_{\rm H2} = 10^5$ cm$^{-3}$. 
In these latter cases the CO(2--1) intensity starts to decrease 
as the line now becomes optically thinner for higher $T_{\rm ex}$. 

Comparison of the model results in Figure \ref{fig9} with the observed line ratios in Figure \ref{fig6} 
therefore gives us an insight on the prevailing physical/chemical conditions of the \ion{C}{1}-enhancement. 
We here only discuss the ratios of the SB galaxies (including the SB ring of NGC 7469) and that of the AGN of NGC 7469
\footnote{Both values from the double Gaussian fit and that of the channel map-basis are discussed in the same manner here.} for simplicity. 
By inspecting Figure \ref{fig9}, it is conceivable that $N_{\rm C^0}/N_{\rm CO} \sim 1$ 
reproduces the ratios of the SB galaxies well, particularly when $T_{\rm kin} <100$ K, irrespective of $n_{\rm H2}$. 
This column density ratio (or [C$^0$]/[CO] abundance ratio) is fully consistent 
with the values measured in previous works for SB galaxies \citep[e.g.,][]{2002A&A...383...82I,2016A&A...592L...3K}. 

The interpretation of the AGN ratios is more complex, which depends on the assumed $n_{\rm H2}$. 
In the cases of $n_{\rm H2} = 10^3$ cm$^{-3}$, we could not find a good solution within the parameter range we searched. 
This in turn suggests that the global gas density of the CND of NGC 7469 is rather high like $\gtrsim 10^4$ cm$^{-4}$. 
In the cases of $n_{\rm H2} = 10^4$ cm$^{-3}$, we need both 
$N_{\rm C^0}/N_{\rm CO} \sim 10$ and $T_{\rm kin} \gtrsim 300-500$ K to explain the observed high ratios. 
This abundance ratio is even $\sim 100\times$ higher than those found in the Milky Way \citep[$\sim 0.1$,][]{2005ApJ...623..889O}, 
and the $T_{\rm kin}$ is also extremely high compared to typical values of Galactic molecular clouds ($\sim 10$ K). 
In the higher $n_{\rm H2}$ cases of 10$^5$ cm$^{-3}$, two possibilities arise. 
One is $N_{\rm C^0}/N_{\rm CO} \sim 3$ with high $T_{\rm kin}$ of $\gtrsim 300-500$ K. 
The other is $N_{\rm C^0}/N_{\rm CO} \sim 10$ with a bit lower $T_{\rm kin}$ of $\sim 100-200$ K. 
In either case, it is required to elevate the [C$^0$]/[CO] ratio by $\sim 3-10\times$ 
and the $T_{\rm kin}$ by $\sim 2\times-10\times$ in the NGC 7469 AGN, as compared to the SB galaxies. 
Therefore, both the prevalent physical and chemical conditions are clearly different 
between these AGN and SB galaxies.

\subsubsection{What causes the CI-enhancement?}\label{sec4.1.3}
We found that the ISM around the AGN of NGC 7469 can be characterized as 
that shows dramatically enhanced $N_{\rm C^0}/N_{\rm CO}$ and temperature 
as compared to those of SB galaxies and molecular clouds in our Galaxy. 
As the unique point of NGC 7469 is obviously the existence of the luminous AGN, 
we should attribute this enhancement to the AGN, or XDR effects. 

In XDRs, X-rays can ionize atoms and molecules directly deeper into the obscuring material, 
which can also cause doubly ionized species for heavier atoms via the Auger mechanism. 
The {\it fast electrons} produced by this primary X-ray ionization further causes, 
secondary ionization, efficient gas heating due to Coulomb interaction, 
as well as photodissociation by internally generating UV photons. 
According to the prescription of \citet{1996ApJ...466..561M}, 
a key parameter to discuss XDR properties is the effective ionization parameter, 
which shapes gas temperature and chemical structures. 
It is expressed as 
\begin{equation}
\xi_{\rm eff} = 1.26 \times 10^{-4} \frac{F_{\rm X}}{n_5 N^\phi_{22}}, 
\end{equation} 
where $F_{\rm X}$ is the incident 1-100 keV flux in units of erg s$^{-1}$ cm$^{-2}$, 
$n_5$ is the gas volume density in units of 10$^5$ cm$^{-3}$, 
$N_{22}$ is the attenuating column density in units of $10^{22}$ cm$^{-2}$, respectively. 
The parameter $\phi$ is related to the photon index of an X-ray SED ($\Gamma$) as $\phi = (\Gamma + 2/3)/(8/3)$. 
This parameter is set to 0.9 (or $\Gamma \sim 1.8$) based on actual X-ray observations of NGC 7469 \citep{2007MNRAS.382..194N}. 
With this $\Gamma$, we also estimate the 1-100 keV luminosity of NGC 7469 as $5.6 \times 10^{43}$ erg s$^{-1}$. 
This X-ray luminosity is comparable to, or even larger than, the {\it upper limit} of the cosmic ray 
($>10^{18}$ eV proton) luminosity of NGC 7469 ($\sim 5 \times 10^{43}$ erg s$^{-1}$ for example, 
for the case of the spectral index of 2.4 and the cutoff energy of 10$^{20.5}$ eV for the injection cosmic ray spectrum) 
measured over the whole galaxy-scale \citep[][]{2013JCAP...12..023S}. 
Therefore, we consider that the X-rays are the prime driver of dissociation/ionization at the CND of NGC 7469. 

Suppose a case of $n_5 = 1$ and $N_{22} = 10$ (typical values for CND-scale gas) for simplicity
\footnote{Indeed, \citet{2014A&A...570A..28V} obtained $n_{\rm H2} \gtrsim 10^{4.5-5}$ cm$^{-3}$ by modeling CO line ratios measured at the CND of NGC 1068.}, 
we obtain $\log \xi_{\rm eff} \sim -2.5$ at a distance of 50 pc from the nucleus: 
we fully covered this area by the fixed 0$\arcsec$.38 aperture. 
According to the one-zone dense ($n_{\rm H2} = 10^5$ cm$^{-3}$) 
XDR chemical model of \citet{1996ApJ...466..561M}, 
we can expect $N_{\rm C^0}/N_{\rm CO} \sim 10$ and $T \sim 300$ K for this $\xi_{\rm eff}$, 
which accords well with the results of our non-LTE analysis. 
Note that, however, there is a drastic increase in $N_{\rm CO}$, 
hence a correspondingly drastic decrease in $N_{\rm C^0}/N_{\rm CO}$, 
toward $\log \xi_{\rm eff} \lesssim -2.5$ according to the model. 
On the other hand, it is also possible that we may be overestimating 
the actual size of the XDR (or the region that dominantly emits [\ion{C}{1}](1--0)) 
as the incident X-ray flux may be attenuated by intercepting ISM 
before reaching a cloud of our interest that is located away from the center. 
A sort of warping of the CND \citep[e.g.,][]{2000ApJ...533..850S} would also be potentially important 
as it easily alter the amount of intercepting ISM, 
although we do not see significant warping in the case of NGC 7469 based on our dynamical modelings (\S~4.2). 
In any case, it is vital to perform further higher resolution observations 
to map the density structure, as well as $\xi_{\rm eff}$ inside this CND to robustly discuss the abundance variation 
\citep[see example high resolution observations toward nearby AGNs to constrain these parameters in,][]{2019PASJ...71...68K,2020ApJ...895..135K}. 
Even so, however, as the XDR models can at least reproduce the observed line ratios 
and physical/chemical conditions we unveiled, 
while PDR models usually do not \citep[e.g.,][]{1999RvMP...71..173H,2005A&A...436..397M}, 
we conclude that there is indeed the influence of the AGN on the surrounding ISM in the form of the XDR.

\subsection{Impact on H$_2$ mass measurements}\label{sec4.2}
As the CND-scale gas of NGC 7469 is characterized by the extreme conditions described above, 
one would have a concern about how it impacts the H$_2$ mass measurements that use CO or C$^0$ lines. 
Indeed, these have been used to measure $M_{\rm H2}$ (or total molecular mass $M_{\rm mol}$) not only in nearby AGNs 
but also in high redshift quasars \citep[e.g.,][]{2011ApJ...730...18W,2020arXiv200603072I}, 
in which we certainly expect the existence of XDRs at their centers. 
For future high resolution observations that directly probe the CND-scale of AGNs at whichever redshift, 
we here try to estimate [\ion{C}{1}](1--0)-to- or CO(1--0)-to-$M_{\rm H2}$ conversion factors in NGC 7469 (CND), 
which will be compared with observationally- or theoretically-derived values known thus far 
\citep[e.g.,][]{2013ARA&A..51..207B,2014MNRAS.440L..81O,2015MNRAS.448.1607G,2017ApJ...840L..18J}. 

For this purpose, we followed the scheme of \citet{2004ApJ...602..148D}. 
First, we decomposed an observed velocity field to obtain 
gas rotation velocity ($V_{\rm rot}$) and dispersion ($\sigma_{\rm disp}$), 
which determine the enclosed dynamical mass ($M_{\rm dyn}$). 
Next, a stellar mass ($M_\star$) profile was modeled based on high resolution {\it HST} maps. 
Then we obtained a total molecular mass as $M_{\rm mol} = M_{\rm dyn}-M_\star-M_{\rm BH}$, which defines CO(1--0) and [\ion{C}{1}](1--0) 
conversion factors ($X_{\rm CO}$ and $X_{\rm CI}$, respectively). 

One big assumption is about the dominant phase of the circumnuclear gas in terms of mass. 
We here assume that H$_2$ still dominates the gas mass budget at the CND of NGC 7469, 
although we found CO and \ion{C}{1} lines are largely affected by the XDR effects, 
which implies that \ion{H}{1} (and likely \ion{H}{2}) contribution can be significant. 
However, we remark that an \ion{H}{1} column density measured by high-resolution 
($0\arcsec.38$) radio absorption line observations toward the center of NGC 7469 
is $N_{\rm H} \sim 4 \times 10^{21}$ cm$^{-2}$ \citep{2002MNRAS.335.1091B}. 
This would not represent a column density toward this type-1 AGN location as the line-of-sight $N_{\rm H}$ 
measured with X-ray observations is $\sim 5 \times 10^{20}$ cm$^{-2}$ \citep[e.g.,][]{2000ApJ...535...58K}. 
Rather, the radio \ion{H}{1} absorption would take place toward bright radio continuum knot(s) 
in the CND (i.e., close to, but not identical to the AGN itself), which are found by Very Long Baseline Interferometry 
(VLBI) observations \citep{2003ApJ...592..804L}. 
Hence the above-mentioned $N_{\rm H}$ of $\sim 4 \times 10^{21}$ cm$^{-2}$ would represent the value at the CND. 
This is significantly smaller than the $N_{\rm H2}$ tentatively inferred from our CO(1--0) observations 
using, for example, the Galactic $X_{\rm CO}$ ($N_{\rm H2} \sim 1.4 \times 10^{23}$ cm$^{-2}$; see footnote 4). 
Given these observational results, we provisionally assume the dominance of H$_2$ 
(or molecular gas) in the mass budget of the region of our interest. 
This assumption, and consequently the conversion factors derived here, 
should be further verified by future high resolution \ion{H}{1} mass measurements.

\begin{figure*}
\begin{center}
\includegraphics[width=\linewidth]{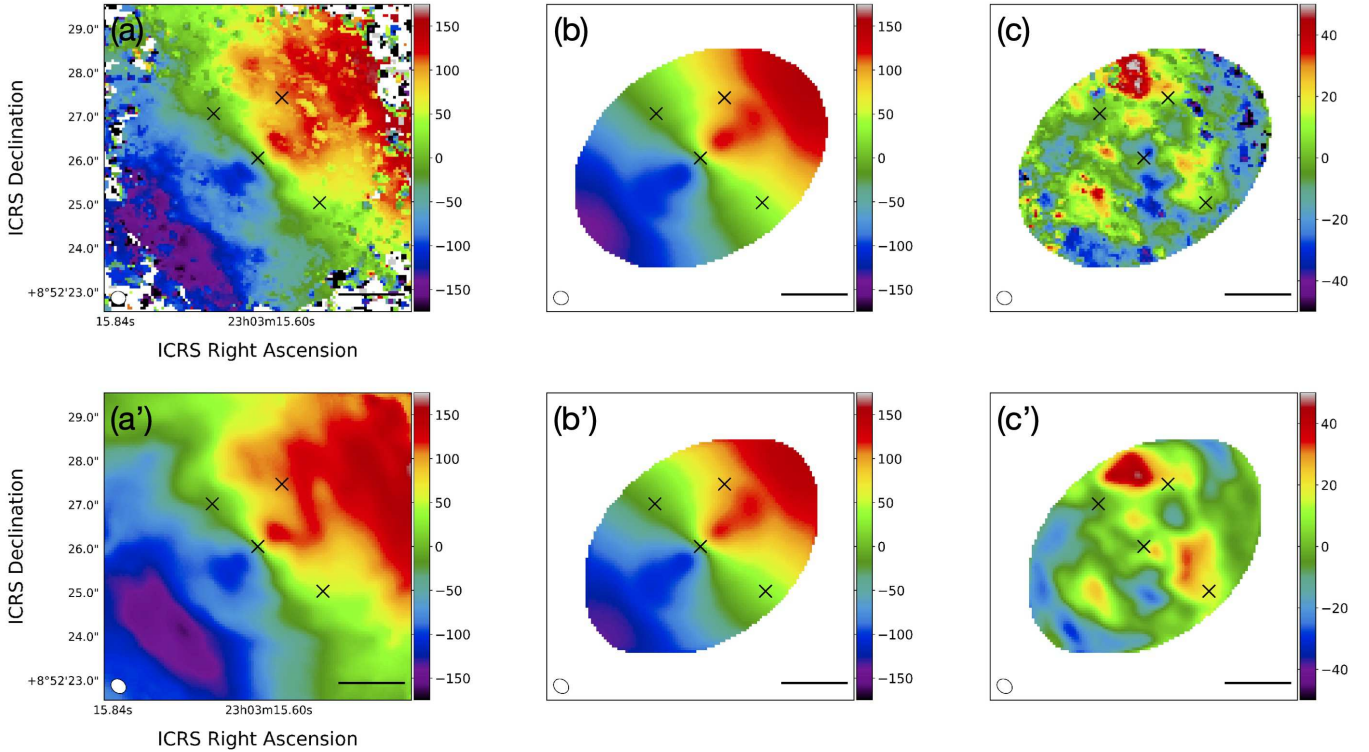}
\caption{
(a) Observed intensity-weighted mean velocity field of the [\ion{C}{1}](1--0) 
emission in the central 7$\arcsec$ ($\sim 2.3$ kpc) of NGC 7469. 
(b) Model velocity field of the [\ion{C}{1}](1--0) by using the tilted-ring method. 
(c) Residual velocity image after subtracting the model from the observed map. 
Residuals are close to 0 km s$^{-1}$ around the AGN. 
(a')(b')(c') Same as the top row, but the cases for the CO(2--1) dynamics. 
In each panel, the four representative positions A--D (Table \ref{tbl6}) are marked by the crosses, 
and the horizontal bar corresponds to 500 pc length. 
}
\label{fig10}
\end{center}
\end{figure*}

In the following, we use the [\ion{C}{1}](1--0) line and the CO(2--1) line cubes, 
both of which have sufficiently high S/N for dynamical modelings. 
Figure \ref{fig10} shows the observed intensity-weighted mean velocity fields of these lines, 
defined as $\langle V \rangle = \Sigma_i S_i V_i/\Sigma_i S_i$ (moment 1) with $3\sigma$ clipping. 
The gas motion is clearly dominated by the galactic rotation with an overall northwest-southeast orientation. 
To extract basic beam-deconvolved dynamical information, we fitted concentric tilted rings 
to the data cubes by using the $^{\rm 3D}$Barolo code \citep{2015MNRAS.451.3021D}. 
The main parameters here are dynamical center, $V_{\rm rot}$, $\sigma_{\rm disp}$, 
radial motion ($V_{\rm rad}$), $V_{\rm sys}$, 
inclination angle ($i$), and position angle (${\rm PA}$), all of which can be varied in each ring. 
However, for a better convergence, we fixed the dynamical center to the AGN position. 
Our initial runs returned $V_{\rm sys}$ fully consistent with our original estimate in \S~3, 
hence we also fixed it to 4920 km s$^{-1}$ (optical convention): 
$V_{\rm rot}$, $\sigma_{\rm disp}$, $V_{\rm rad}$, $i$, and ${\rm PA}$ are thus the major parameters to fit. 
For initial guesses, we set $i = 45\arcdeg$ and ${\rm PA} = 128\arcdeg$ 
based on the previous CO-based dynamical work
\footnote{Note that the Barolo code defines PA as that of the receding half 
of the galaxy taken anticlockwise from the north direction on the sky. 
We thus need to add another 180$\arcdeg$ to the observed PA (i.e., 308$\arcdeg$), which should be put into the code. 
The northern part of the galaxy is the near side to us.} \citep{2004ApJ...602..148D}. 
We modeled 50 concentric rings with $\Delta r = 0\arcsec.05$ starting from $r = 0\arcsec.10$. 
The fitting was evaluated by minimizing the residual amplitude, $|$model--observed data$|$.

\begin{figure}
\begin{center}
\includegraphics[width=\linewidth]{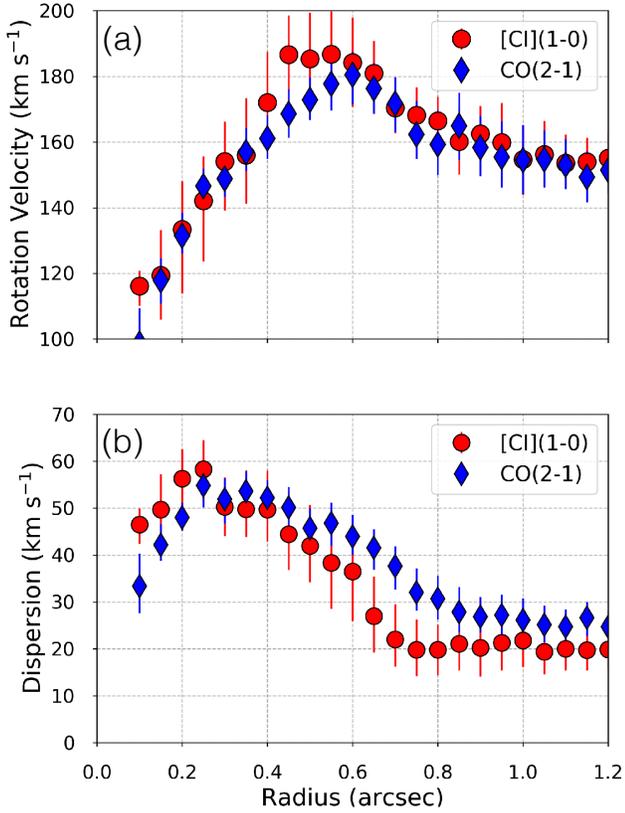}
\caption{
Radial profiles of the beam-deconvolved (a) rotation velocity ($V_{\rm rot}$) 
and (b) velocity dispersion ($\sigma_{\rm disp}$) at the innermost $r < 1\arcsec.2$. 
The cases of the [\ion{C}{1}](1--0) (red circles) 
and the CO(2--1) (blue diamonds) dynamics are displayed. 
}
\label{fig11}
\end{center}
\end{figure}

\begin{figure}
\begin{center}
\includegraphics[width=\linewidth]{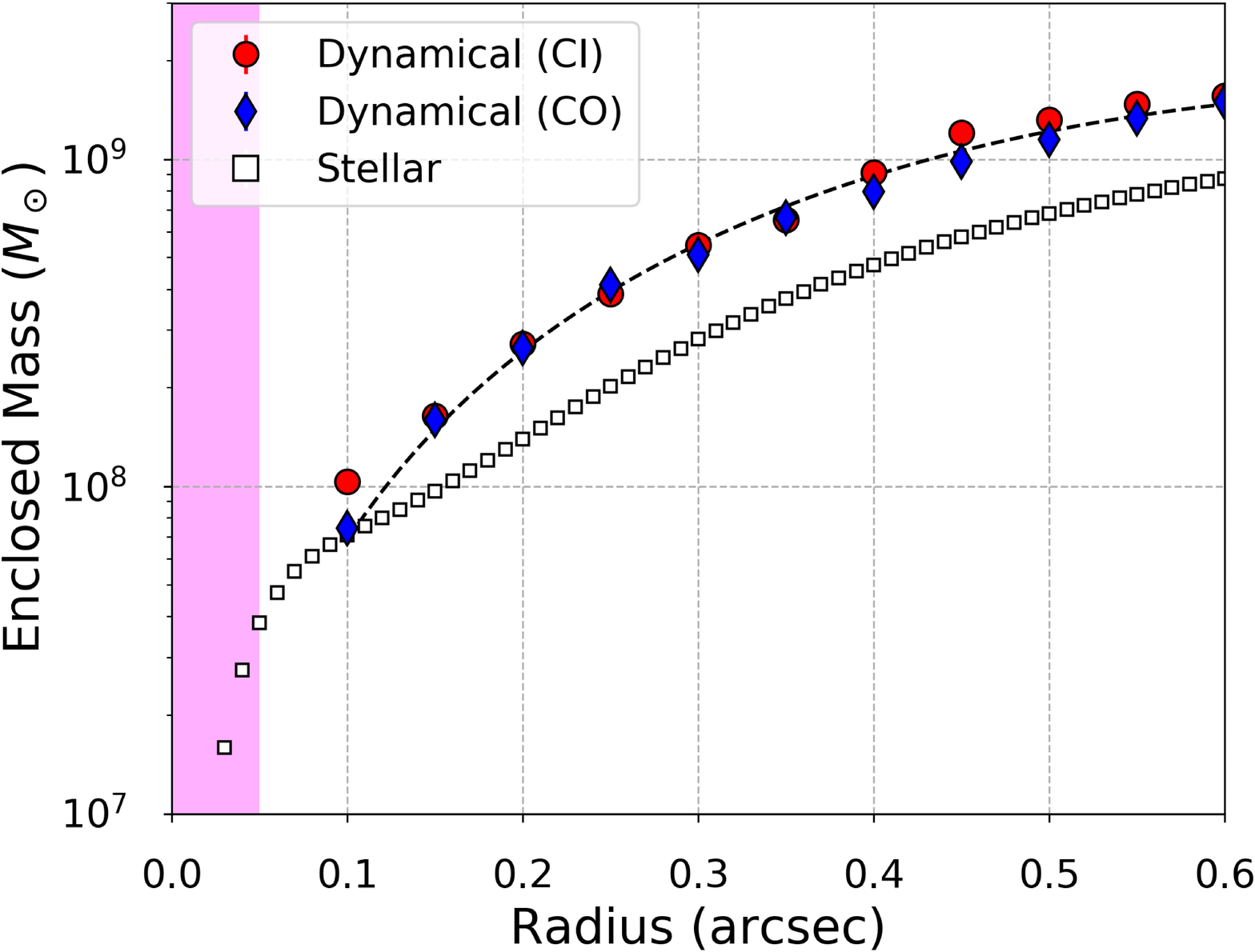}
\caption{
Enclosed stellar mass ($M_\star$) and dynamical mass 
($M_{\rm dyn}$; based on the [\ion{C}{1}](1--0) dynamics or CO(2--1) dynamics) around the center of NGC 7469. 
The $M_\star$ values are returned from the MGE model 
(see details in D. Nguyen et al. in preparation), 
by using the high resolution {\it HST} WFC3/UVIS F547M 
and ACS/WFC F814W maps with the color--$M/L_V$ relation \citep{2001ApJ...550..212B}. 
The region masked out for the $M_\star$ measurement is shaded ($r \leq 0\arcsec.06$). 
Our tentative fit for the $M_{\rm dyn}$ profile by using a 5th order polynomial function is shown by the dashed line. 
}
\label{fig12}
\end{center}
\end{figure}

The modeled mean velocity fields, as well as the residual images after subtracting the models 
from the observed images, are also shown in Figure \ref{fig10}. 
Most of the residual components are minor with $\lesssim 20$ km s$^{-1}$ 
over the modeled region, which manifests the goodness of our fit. 
Figure \ref{fig11} shows the radial profiles of the decomposed $V_{\rm rot}$ and $\sigma_{\rm disp}$. 
Both lines show comparable values within $\sim 15$ km s$^{-1}$ difference, 
suggesting that these trace essentially the same gas rotation in the currently observed regions. 
The variations in $i$ and ${\rm PA}$ are very small 
(within 5$\arcdeg$ and 10$\arcdeg$, respectively) around our initial guesses. 

On the other hand, we found a non-negligible difference in $V_{\rm rad}$: 
it is within $\pm 25$ km s$^{-1}$ for the case of [\ion{C}{1}](1--0) over the all radii, 
while it decreases down to $< -50$ km s$^{-1}$ (inflow) at the innermost five rings for the case of CO(2--1). 
However, we claim that a significant fraction of this $V_{\rm rad}$ is an artifact 
due to the configuration of the CO(2--1) emission distribution around the center, and not due to genuinely that fast inflows. 
The two bright CO(2--1) knots appear at lower and higher velocity than $V_{\rm sys}$, 
spatially at south-west and north-east side of the AGN almost along the minor axis of this galaxy (Figure \ref{fig5}). 
Owing to this chance spatial coincidence of high- and low-velocity bright knots with the minor axis, 
the simple tilted-ring scheme misunderstands this configuration as that caused by fast radial flows. 
The real $V_{\rm rad}$ would be much milder such as seen in the [\ion{C}{1}](1--0) data. 

Another notable feature is an upturn in $V_{\rm rot}$ from $r \sim 1\arcsec.0$ to $\sim 0\arcsec.5$, 
which can be regarded as a sign of the Keplerian motion due to the central SMBH. 
Note that the sphere of influence (SOI) radius should be $\sim 3$ pc or $\sim 0\arcsec.01$ 
for the case of NGC 7469 with $M_{\rm BH} = 1.06 \times 10^7~M_\odot$ \citep{2014ApJ...795..149P} 
and the stellar velocity dispersion of the bulge of $\sim 152$ km s$^{-1}$ \citep{2004ApJ...615..645O}, 
which is much smaller than our beam size. 
However, this SOI criterion does not necessarily apply for $M_{\rm BH}$ measurements 
using gas-dynamical method as shown in previous works 
\citep[e.g.,][]{2014MNRAS.443..911D,2019ApJ...872..104N,2019arXiv190203813N}. 
Further detailed dynamical modeling including the Markov Chain Monte Carlo method and the Bayesian inference 
to derive $M_{\rm BH}$ of NGC 7469 will be presented in D. Nguyen et al. (in preparation). 

With the decomposed value of $V_{\rm rot}$, we compute $M_{\rm dyn}$ as 
\begin{equation}\label{eq8}
\begin{split}
M_{\rm dyn} &= \frac{r V^2_{\rm rot}}{G}  \\
&= 230 \left( \frac{r}{\rm pc} \right) \left( \frac{V^2_{\rm rot}}{\rm km~s^{-1}} \right), 
\end{split}
\end{equation}
where $G$ is the gravitational constant. 
The uncertainty of $V_{\rm rot}$ is typically $\sim 10\%$, 
which propagates to the uncertainty of $M_{\rm dyn}$. 
The resultant $M_{\rm dyn}$ values are shown in Figure \ref{fig12}: 
we find very consistent values between the [\ion{C}{1}](1--0)-based and the CO(2--1)-based $M_{\rm dyn}$. 
The derived $M_{\rm dyn}$ of the concentric rings are further interpolated by 
a fifth-order polynomial function to estimate values at a given radius. 
Consequently, within $r = 0\arcsec.19$ or $\theta = 0\arcsec.38$ region where we took line ratios (\S~3.3), 
we find $M_{\rm dyn} = 2.3 \times 10^8~M_\odot$ after averaging [\ion{C}{1}](1--0)-based 
and CO(2--1)-based values, which has $\sim 15\%$ uncertainty. 

As the next step, we used {\it HST} WFC3/UVIS F547M map and ACS/WFC F814W map to estimate an $M_\star$ profile. 
Details of the {\it HST} data analysis and $M_\star$ measurements 
will also be presented in D. Nguyen et al. (in preparation). 
The astrometry of these {\it HST} data were corrected by using the {\it Gaia} coordinates of NGC 7469. 
Here we used an empirical relation between the stellar continuum color and the mass-to-light ratio ($M/L$) 
developed by \citet{2001ApJ...550..212B} to measure $M_\star$. 
We assumed F547M $\sim V$ band and F814W $\sim I$ band, respectively, 
and applied the $(V-I)$-to-$M/L_V$ relation of that work. 
This procedure was performed in concentric elliptical annuli with the multiple Gaussian expansion model 
\citep[MGE,][]{1994A&A...285..723E,2002MNRAS.333..400C}. 
The Gaussians of the MGE model are then deprojected analytically with their specific axes ratios 
(i.e., the ratio of the semiminor axis to the semimajor axis of each concentric elliptical Gaussian) 
to reconstruct a three-dimensional mass distribution and calculate the enclosed mass profile (Figure \ref{fig12}). 
Note that we masked the bright central AGN ($r \leq 0\arcsec.06$; 
comparable to the FWHM of the PSF of the {\it HST} data = $0\arcsec.08$) that saturates the F814W map 
at that position as well as the SB ring for our modeling. 
The enclosed $M_\star$ within the $r = 0\arcsec.19$ is $1.3 \times 10^8~M_\odot$. 
We found that the color variation is significant ($\sim 0.2$ mag) 
at $r < 10\arcsec$ of NGC 7469 due likely to the complex stellar population and dust extinction, 
which imposes a large uncertainty on our $M_\star$ measurement as $\sim 0.2$ dex. 
This would dominate the total uncertainty in our conversion factors. 

By subtracting the $M_\star$ and $M_{\rm BH}$ from $M_{\rm dyn}$, 
we derive the total gas mass at $r \leq 0\arcsec.19$ as $M_{\rm mol} = 9.5 \times 10^7~M_\odot$. 
The molecular hydrogen mass is $M_{\rm H2} = 6.7 \times 10^7~M_\odot$ 
for the fractional abundance of hydrogen nuclei of 71\%. 
Note that we do not consider the contribution of dark matter here 
as we focus only on the very central region of a galaxy. 
The CO(1--0) and [\ion{C}{1}](1--0) line fluxes and luminosities measured over that area are, 
1205 K km s$^{-1}$ and $2.3 \times 10^7$ K km s$^{-1}$ pc$^2$ for CO(1--0), 
as well as 1114 K km s$^{-1}$ and $2.1 \times 10^7$ K km s$^{-1}$ pc$^2$ 
for [\ion{C}{1}](1--0), respectively (Table \ref{tbl7}). 
Therefore, the conversion factors to the total molecular mass are, 
$\alpha_{\rm CO}$ = 4.1 $M_\odot$ (K km s$^{-1}$ pc$^2$)$^{-1}$ 
and $\alpha_{\rm CI}$ = 4.4 $M_\odot$ (K km s$^{-1}$ pc$^2$)$^{-1}$, respectively. 
The corresponding factors for $N_{\rm H2}$ measurements are, 
$X_{\rm CO} = 1.9 \times 10^{20}$ cm$^{-2}$ (K km s$^{-1}$)$^{-1}$ 
and $X_{\rm CI} = 2.1 \times 10^{20}$ cm$^{-2}$ (K km s$^{-1}$)$^{-1}$, respectively. 
The uncertainties of $M_{\rm dyn}$ measurement and $M_\star$ measurements jointly yield the uncertainties of these factors, 
which are then $\sim 0.3$ dex, although it is hard to constrain this number accurately. 

Our $X_{\rm CO}$ is very comparable to the canonical Milky Way value \citep{2013ARA&A..51..207B,2014MNRAS.440L..81O}. 
It is $\sim 3\times$ larger than the typical value inferred for AGNs 
based on kpc-scale resolution observations \citep[e.g.,][]{2013ApJ...777....5S}, 
and is also $1.6\times$ larger than the $X_{\rm CO}$ previously measured for NGC 7469 
itself over the central $\sim 1.7$ kpc region \citep{2004ApJ...602..148D}. 
A smaller $X_{\rm CO}$ than the Milky Way value has been observed in nuclear regions of normal spiral galaxies as well, 
likely due to tidal effects on molecular clouds \citep{2004AJ....127.2069M,2008ApJ...675..281M}. 
Hence, our $X_{\rm CO}$ seems contradictory to the previously reported trend 
of decreasing $X_{\rm CO}$ toward more active environments including AGNs. 
This is due to the high spatial resolution of this work that allows us to probe the XDR 
(i.e., the region at which CO molecules are dissociated to some level) of NGC 7469 properly, 
which should result in a large $X_{\rm CO}$ for a given H$_2$ mass. 
If this is the case, the consistency of our $X_{\rm CO}$ 
to the Milky Way value is simply a chance coincidence, 
as the underlying physical/chemical conditions must be different greatly. 
It is noteworthy in this context that our relatively large $X_{\rm CO}$ is consistent 
with the value reported by \citet{2018ApJ...852...88W}, 
who simulated CND-scale CO properties around an AGN 
by incorporating the XDR chemical network of \citet{2005A&A...436..397M}. 
However, \citet{2018ApJ...852...88W} also claimed that there can be a quite large 
dispersion in $X_{\rm CO}$ up to one order of magnitude. 
Therefore, similar high resolution observations toward a statistical number of AGNs 
are required to assess $X_{\rm CO}$ and its scatter at their close vicinities. 

Our C$^0$ conversion factors ($\alpha_{\rm CI}$ and $X_{\rm CI}$) are $\sim 5\times$ 
smaller than the values expected for Galactic star-forming clouds \citep{2014MNRAS.440L..81O,2015MNRAS.448.1607G}. 
\citet{2017ApJ...840L..18J} estimated $\alpha_{\rm CI} = 7.6$ $M_\odot$ (K km s$^{-1}$ pc$^2$)$^{-1}$ 
for a sample of 71 nearby U/LIRGs based on {\it Herschel} observations. 
Our conversion factors are still $\sim 2\times$ smaller than this value, 
as well as than that individually derived for another ULIRG NGC 6240 \citep{2018ApJ...863..143C}. 
The observational estimation of $X_{\rm CI}$ critically depends on the inverse of the C$^0$ relative abundance to H$_2$: 
for example \citet{2017ApJ...840L..18J} assumed [C$^0$]/[H$_2$] = $3 \times 10^{-5}$, which is a value 
of the $z = 2.5$ Cloverleaf quasar estimated over its galaxy-scale \citep{2003A&A...409L..41W}. 
If the [C$^0$]/[H$_2$] value for the central $\sim 100$ pc region of NGC 7469 is larger 
than the values of our Galactic star-forming clouds and those of U/LIRGs, 
our smaller conversion factors are understandable. 
This is again very likely the case as we now probe the XDR of NGC 7469, 
where we can expect a ratio as high as $\gtrsim 10^{-4}$ according to XDR models \citep[e.g.,][]{1996ApJ...466..561M}. 

As our method to estimate the conversion factors is rather direct, 
they would be useful to estimate $M_{\rm H2}$ at the vicinities of AGNs. 
However, as the nature of an XDR is dependent on several physical parameters including X-ray luminosity, 
gas density, and attenuating column density, similar efforts to what we have performed here 
are highly required to obtain characteristic conversion factors, if exist, and to assess their scatter.

\section{Summary}\label{Sec5}
In this paper we present high resolution ($\sim 130$ pc) ALMA observations 
of multiple CO and C$^0$ lines toward the central kpc region of the luminous type-1 Seyfert galaxy NGC 7469. 
The region consists of the CND (central $\sim 1\arcsec$) and the surrounding SB ring (radius $\sim 1\arcsec.5$). 
All of the targeted emission lines, namely CO(1--0), CO(2--1), CO(3--2), $^{13}$CO(2--1), and [\ion{C}{1}](1--0), 
are successfully detected both in the CND and the SB ring. 
Thanks to the high resolution, we could reliably measure the line fluxes and their ratios, 
which are used to discuss the nature of ISM particularly at the vicinity of the AGN in a context of X-ray dominated region (XDR) chemistry. 
Our findings of this work are summarized in the following. 

\begin{itemize}
\item[1.] $^{12}$CO lines are bright both at the CND and the SB ring, which defines the base gas distribution of this galaxy. 
On the other hand, $^{13}$CO(2--1) is very faint at the CND, while [\ion{C}{1}](1--0) emission is concentrated toward the CND. 
The [\ion{C}{1}](1--0) emission distribution clearly peaks at the exact AGN position, whereas CO and $^{13}$CO distributions do not. 
This is unlikely due to absorption effect considering the type-1 Seyfert geometry, 
and already hints at the influence of the AGN on the gas physical/chemical structures. 
Given the centrally-peaked distribution, the [\ion{C}{1}](1--0) emission 
also defines the systemic velocity of this galaxy as $V_{\rm sys} = 4920$ km s$^{-1}$. 
\item[2.] Consequently, we found the line flux ratios of [\ion{C}{1}](1--0)/CO(2--1) ($\equiv R_{\rm CI/CO}$) 
and [\ion{C}{1}](1--0)/$^{13}$CO(2--1) ($\equiv R_{\rm CI/13CO}$), measured over $\sim 130$ pc area, 
are dramatically different between the CND and the SB ring (i.e., \ion{C}{1}-enhancement in the CND). 
There is a trend of increasing these ratios in AGNs as compared to SB galaxies or quiescent galaxies 
as found in the compilation of the single dish-based data. 
But the ratios we revealed at the CND of NGC 7469 are extraordinary 
($R_{\rm CI/CO} \sim 0.5$ and $R_{\rm CI/13CO} \sim 20$ as integrated flux ratios, 
or $R_{\rm CI/CO} \sim 0.8$ and $R_{\rm CI/13CO} \sim 25$ as channel map-based ratios), 
which have never been observed at the spatial scales probed here: 
these AGN ratios are $\sim 3\times$ and $\sim 10\times$ higher than typical values of SB galaxies. 
\item[3.] The high ratios observed at the CND indicate the power of the high resolution provided by ALMA, 
which allows us to selectively probe the regions influenced by the AGN (or likely XDR). 
We suggest that these ratios would have a potential as a submm diagnostic method of the underlying heating sources (AGN vs SB). 
\item[4.] Our LTE and non-LTE analysis of the line ratios both indicate that we need an elevated C$^0$/CO abundance ratio 
around the AGN as compared to that of the SB ring of this galaxy, likely by $\sim 3-10\times$, 
to reproduce the different line ratios observed. 
Moreover, we need a higher gas kinetic temperature ($\sim 100-500$ K) 
as well around the AGN than those at the SB ring ($<100$ K). 
Note that within the parameter range studied here, both [\ion{C}{1}](1--0) and $^{13}$CO(2--1) lines are optically thin in most cases. 
\item[5.] The unique abundance ratios and high gas temperature are well in accord with the scenario 
that the AGN influences the surrounding physical and chemical structures of the ISM in the form of XDR. 
\item[6.] We modeled the velocity fields of the [\ion{C}{1}](1--0) line and the CO(2--1) line cubes by using a tilted-ring method. 
We found consistent rotation velocity ($V_{\rm rot}$) and dispersion ($\sigma_{\rm disp}$) between these two lines. 
Using the $V_{\rm rot}$, we could measure enclosed dynamical mass ($M_{\rm dyn}$) inside a given radius. 
For example, $M_{\rm dyn}$ at $r \leq 0\arcsec.19$ ($\sim 60$ pc) is $2.3 \times 10^8~M_\odot$. 
\item[7.] As we revealed unusual ISM conditions around the AGN of NGC 7469, 
we computed dedicated conversion factors from CO(1--0) and [\ion{C}{1}](1--0) 
luminosities to the total molecular (or H$_2$) mass 
by also using the results of our dynamical modelings. 
Our big assumption is the dominance of H$_2$ gas over the gas mass budget 
at the innermost $\sim 100$ pc region of NGC 7469. 
We obtained $\alpha_{\rm CO} = 4.1~M_\odot$ (K km s$^{-1}$ pc$^2$)$^{-1}$ and 
$\alpha_{\rm CI} = 4.4~M_\odot$ (K km s$^{-1}$ pc$^2$)$^{-1}$ for the central $\sim 100$ pc of NGC 7469. 
Alternatively $X_{\rm CO} = 1.9 \times 10^{20}$ cm$^{-2}$ (K km s$^{-1}$)$^{-1}$ 
and $X_{\rm CI} = 2.1 \times 10^{20}$ cm$^{-2}$ (K km s$^{-1}$)$^{-1}$. 
The [\ion{C}{1}](1--0) conversion factors of NGC 7469 are smaller than those derived for Galactic star-forming regions and for nearby U/LIRGs, 
which would be a natural consequence of elevated C$^0$ abundance in the XDR of NGC 7469. 
\end{itemize}

The \ion{C}{1}-enhancement captured in this work is a quite dramatic phenomenon. 
As it is only based on the results of NGC 7469, we will further investigate the trend by increasing the sample galaxies, 
and will extensively discuss the ISM properties by also comparing the ratios with state-of-the-art chemical models. 
For the particular case of NGC 7469, we guide readers to our forthcoming paper (S. Nakano et al. in preparation) for such comparisons. 
If the trend is confirmed, the \ion{C}{1}-enhancement will be used as a submm energy diagnostic method 
that is applicable to dusty environments as this wavelength does not suffer from severe dust extinction 
\citep[except for some very extremely dusty cases seen in so-called compact obscured nuclei = CONs, e.g.,][]{2013ApJ...764...42S,2019A&A...627A.147A}, 
as well as to high redshift galaxies owing to the high $\nu_{\rm rest}$ of [\ion{C}{1}](1--0).

\acknowledgments 
We thank the anonymous referee for her/his thorough 
reading and very constructive feedback, which improved this work greatly. 
This paper makes use of the following ALMA data: 
ADS/JAO.ALMA\#2017.1.00078.S. 
ALMA is a partnership of ESO (representing its member states), 
NSF (USA) and NINS (Japan), together with NRC (Canada), 
MOST and ASIAA (Taiwan), and KASI (Republic of Korea), 
in cooperation with the Republic of Chile. 
The Joint ALMA Observatory is operated by ESO, AUI/NRAO and NAOJ. 
We used data based on observations with the NASA/ESA {\it Hubble Space Telescope} 
and obtained from the Hubble Legacy Archive, which is a collaboration between 
the Space Telescope Science Institute (STScI/NASA), 
the Space Telescope European Coordinating Facility (ST-ECF/ESA), 
and the Canadian Astronomy Data Centre (CADC/NRC/CSA). 
T.I., S.B, T.K. and K.K are supported by Japan Society for the Promotion of Science (JSPS) 
KAKENHI Grant Number JP20K14531, 19J00892, JP20K14529, and JP17H06130, respectively. 
K.K. is also supported by the NAOJ ALMA Scientific Research Grant Number 2017-06B. 
T.I. was supported by the ALMA Japan Research Grant of NAOJ ALMA Project, NAOJ-ALMA-241. 

\bibliography{ref}

\end{document}